# Substrate effect and nanoindentation fracture toughness based on pile up and failure

Arnab S. Bhattacharyya*, R. Praveen Kumar,
Rohit Mandal, Nikhil Kumar, N. Rajak, Abhishek Sharma, Shashi Kant

Centre for Nanotechnology
Central University of Jharkhand, Brambe, Ranchi: 835205, India

2006asb@gmail.com, arnab.bhattacharya@cuj.ac.in

## Abstract

The effect of substrate was studied using nanoindentation on thin films. Soft films on hard substrate showed more pile up than usual which was attributed to the dislocation pile up at the film substrate interface. The effect of tip blunting on the load depth and hardness plots of nanoindentation was shown. The experimental date of variation of Vickers hardness with film thickness and loads were fitted and new parameters were analyzed. The delaminated area was analyzed using geometrical shapes using optical view of the failure region along with the load displacement Indentation fracture using Nanoindentation using Berkovich indenter has been studied. Indentation fracture toughness ($K_R$) was analyzed based on computational programs. The contact mechanics during nanoindentation was studied with parameters related to indenter shape and tip sharpness. Elastic, plastic and total energies were computationally determined. The energy difference was related to shear stress being generated with elastic to plastic transition. Change in the nature of residual stress was related to film thickness.

Keywords: substrate effect, nanoindentation, film failure, indentation fracture toughness

# 1 Introduction

Hardness is an important mechanical property of materials required in research and industry. It basically refers to resistance imposed to localized plastic deformation. The deformation is done by indentation. The indentation process can be done at both micro and nano level. The ratio of load applied to indentation area is a measure of the hardness. The shape of the indentation varies with the geometry of the indenter used [1, 2]. Determination of hardness of thin films is challenging as substrate effect comes into the scenario. Material scientists have now and then come up with new ideas to minimize the substrate effect and obtain only the film hardness. But the task is complicated as technologically the whole system including the film and substrate is useful the mechanical property of the film and substrate compliments each other. There are lot of models and ways that have been proposed to determine the composite hardness or to identify the substrate effect in thin film/substrate systems which can be considered as super sufficient. Finite element models have also been proposed to understand the substrate effect. In this research work, we have studied the substrate effect during indentation and tried to establish a new approach based on experimental curve fitting and simulation.

# 2 Nanoindentation

Nanoindentation is an effective tool for determining mechanical properties at nanoscale. The load-depth curve obtained during nanoindentation gives the load which is required to penetrate a certain depth in the material and also provides the elastic and plastic work done during the indentation process. The area inside the loading and unloading curves is the plastic area. For a 100 % elastic contact, the unloading curve will retrace exactly the same curve followed during the loading process and the effective area will be zero. However as the contact becomes more and more plastic the unloading curve starts to deviate from the loading curve and the area between the two curves increases. For considerably ductile materials like metals, the plastic area is much larger compared to hard materials like ceramics. In case of thin films however the situation is complicated as there is a major influence from the substrate with increase in depth of penetration. Nanoindentation is specifically useful for thin films where a shallow depth of indentation is used. Usually it has been seen that there is no substrate effect up to 10 % of the film thickness. However the matter is not also not as effective for very thin films as it seems (discussed later). The lateral resolution is also taken as the seven times the contact depth for Berkovich pyramial indenter [3-8]. The schematic diagram of a Berkovich indenter is shown below (Fig 1).

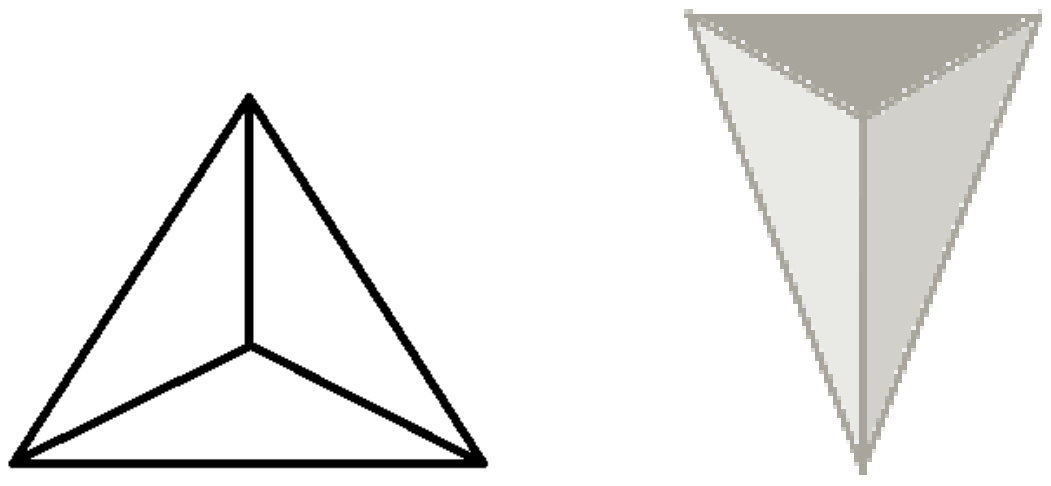
(a) (b)

Fig 1 : Top and side veiw of a three sided pyramidal Berkovich indenter

During nanoindentation a Berkovich indenter with 70.3° effective cone is angle pushed into the material and withdrawn and the hardness value is obtained from the ratio of indentation load and projected contact area (*eqn 1*). where $P_{max}$ is the maximum load applied and $A_{projected}$ is the corresponding projected contact area [5 – 8]. The projected area is estimated from an empirical function involving the contact depth *(*$h_c$*)* given by *eqn.2* A typical load depth curve and corresponding hardness plot is shown in fig 2 .

Hardness and modulus are determined from nanoindentation load depth curve using the standard Oliver and Pharr method. Some modification of the Oliver and Pharr method by intrducing nw parameters have also been reported[11, 12]. The Oliver and Pharr method however is not efficent eneough when pile up or sink-in occurs in the specimen. Pile up causes an understimetion of the contact area and hence an over estimation of hardness[13].

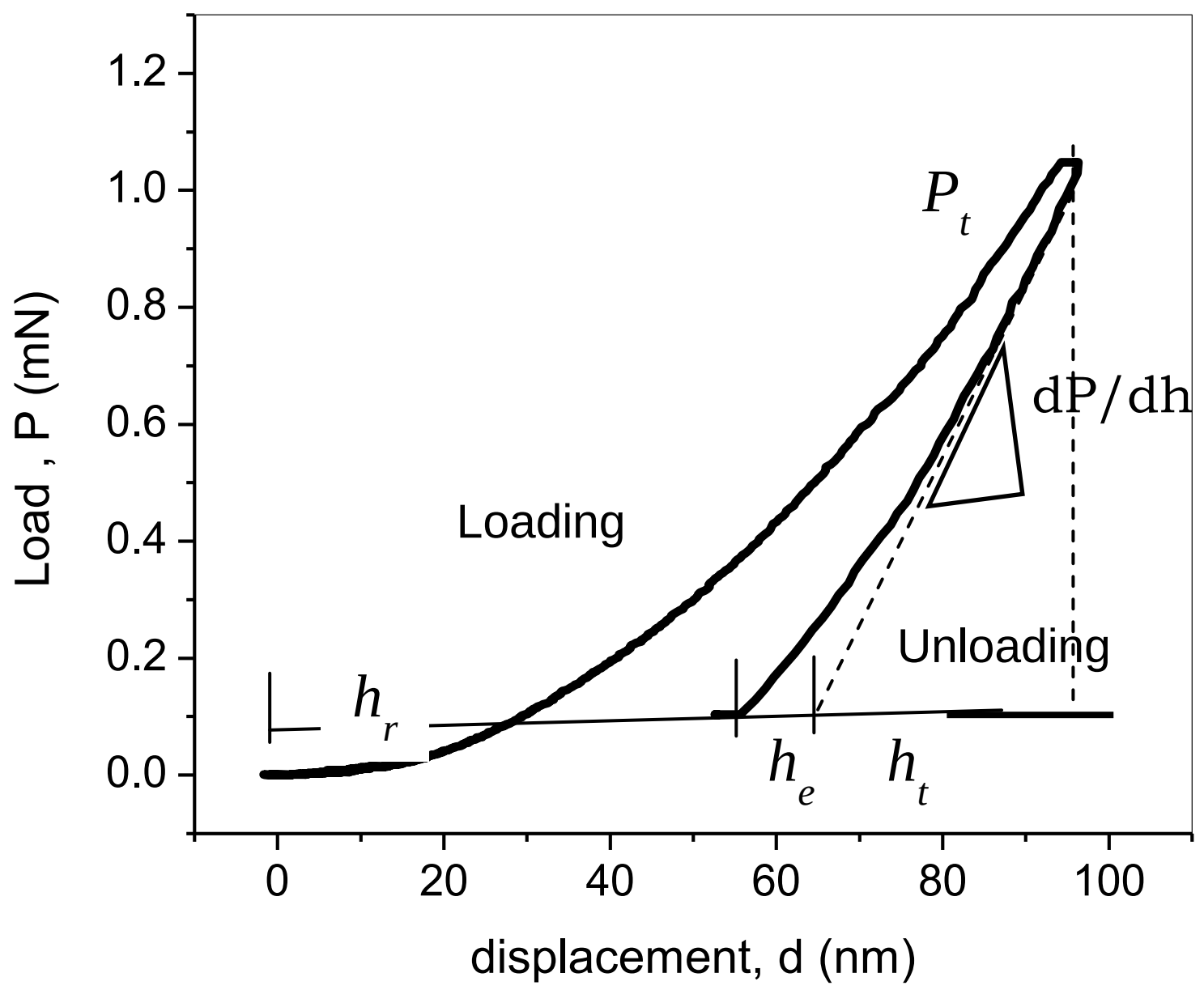

Fig 2 : Load-depth curve during nanoindentation

$$H = \frac{P_{\max}}{A_{projected}} \qquad (1)$$

$$A = 24.5 h_c^{\ 2} \qquad (2)$$

The composite hardness, $H_c$ of a film substrate system is usually expressed as in eqn 3 where $H_f$ is the film hardness, $H_s$ the substrate hardness and $\varphi_H$ a factor depnding upon the model used.

$$H_c = H_s + (H_f - H_s)\varphi_H \qquad (3)$$

# 3 Pile-up and sink-in

Pile up is observed for softer coatings on harder subjects like metals Al, Cu and Au deposited on Si which has a lot of technological importance especially in semiconductor industry. The amount of pile up occuring is again measured through ratio of corner area and actual contact area of the indenter impression ( $A_{actual}$ / $A_{cc}$) as depicted in fig 3 and 4

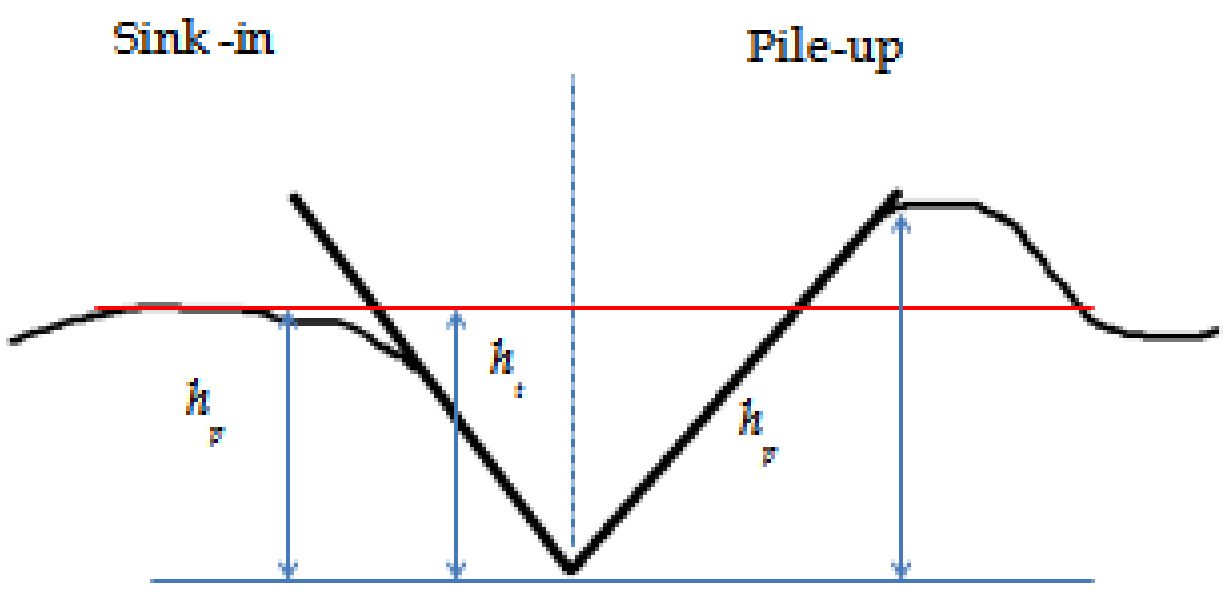

$h_r = h_p - h_t$ =residual depth
$h_r / h_t$ extent of pile-up

Fig 3 : Schematic diagram of Sink-in and pile-up during indentation

A value of ratio more than one indicates imlies file pile up whereas less than one imples sink in effect. Both pile up and sink in are phenomena associated with plastic deformation and their degree depends on E/Y ratio and strain hardening properites of the material. For materials undergoing less strain hardening with high E/Y ratio most of the plastic deformation occurs near the indenter which results in piling up. For materials having low value of E/Y (eg: glasses and ceramics) the plastic zone is typically confined within the boundary of the circle of contact and the elastic deformations that accomadate the volume of the indentation are spead out at a greater distance from the indenter.As a result sink-in in more likely to occur. Also materials showing strain hardening , the plastic zone becoems harder as the amount of deformation increases making the outermosr most material in the plastic zone softer is more susceptible to plastic deformaton as the indentation proceeds and the materials is driven deeper into the specimen material causing the sink–in.

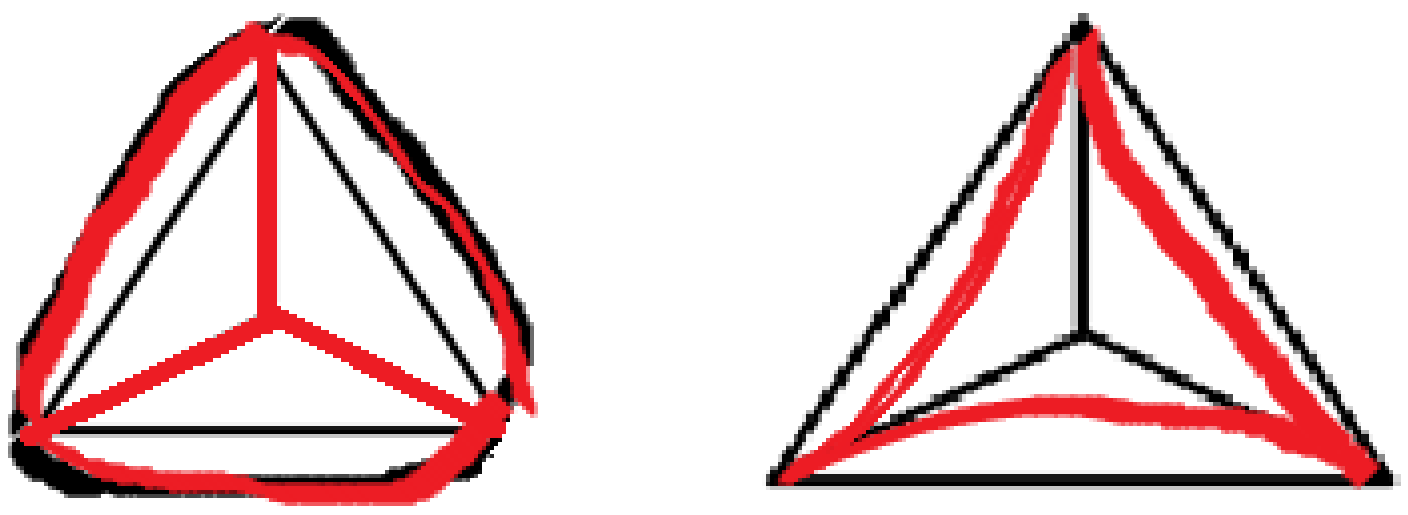

Fig 4: Actual and corner to corner area during indentation pile up

The size of the plastic volume in the film and the substrate determines the hardness. The contact radius to film thickness ratio is directly proportional to the constraint on the film under indentation. The constraint is offered by the substrate on the plastic flow of the film for a softer film.

Different values of hardness for harder film on softer substrates are obtained at different depths for different indentation process although the films were of similar thickness. This was mainly due to nanoindentation tip wear. Bull has shown the effect of tip where it is observed that for the same load the penetration is much lower for a blunt tip and lesser amount of plastic work is done (curve area)[9].We have on the other hand have shown the same thing by keeping the depth of penetration same in both the case (~500nm). It can be observed that much higher load is required to attain the same penetration depth for a blunt tip compared to a sharp tip. The hardness plots show that although nanoindentation will always give the same hardness for a particular film, the depth of attaining film hardness may alter depending upon the tip wear (Fig 5). However it is advisable to use a sharper tip for hardness or to calibrate the instrument for tip correction. The hardness plots for the optimized conditions were performed using a sharper tip providing the required thickness at much lower loads.

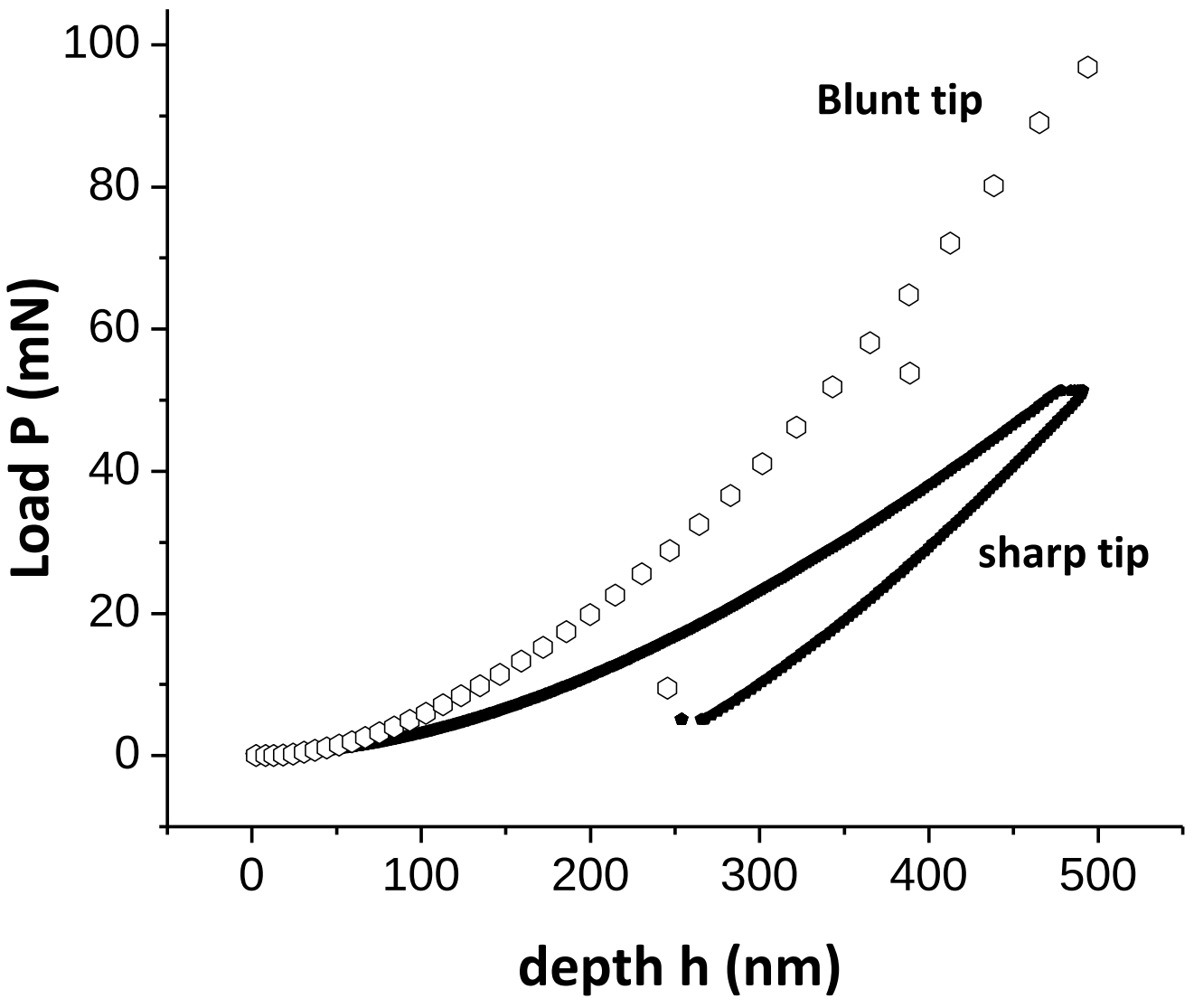

*(a)*

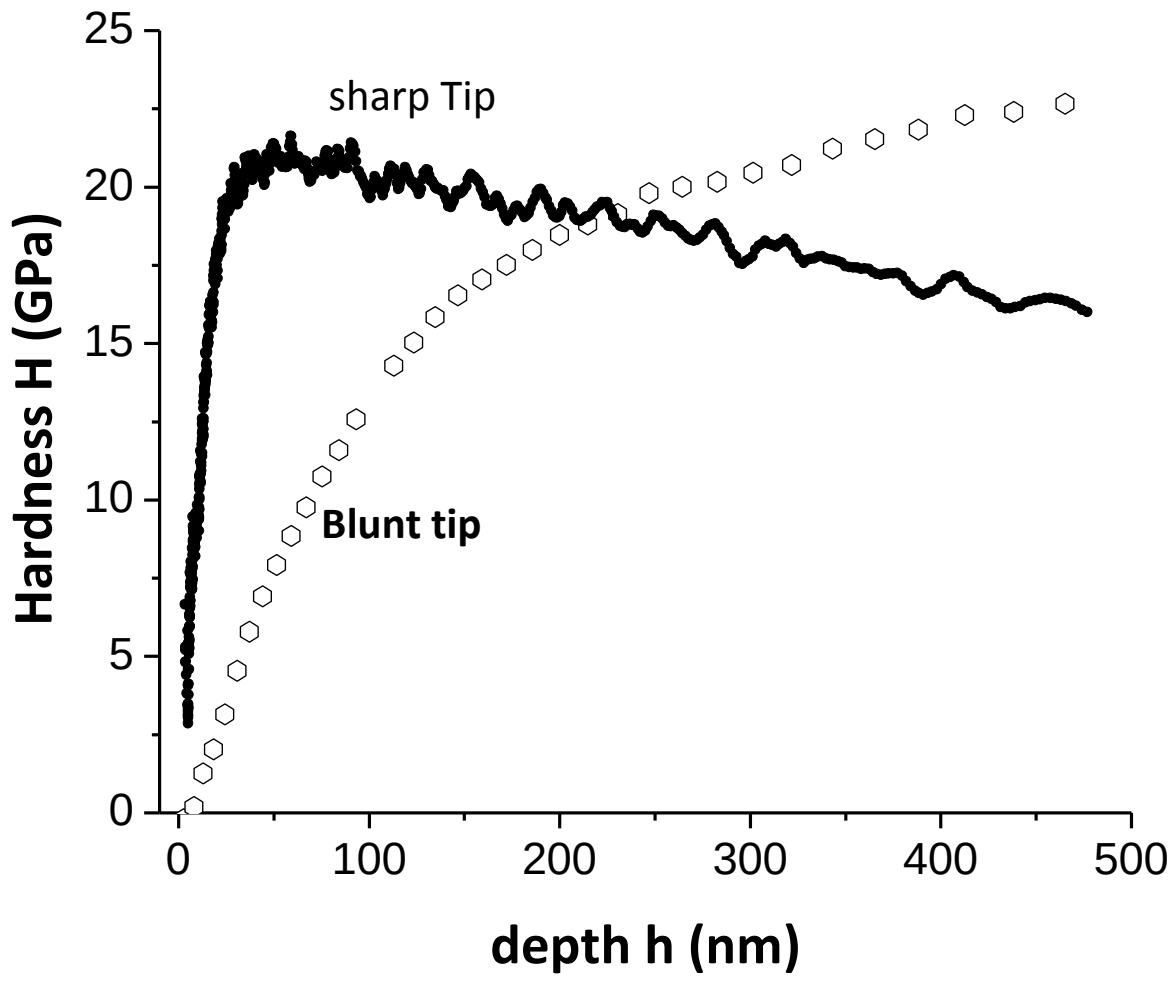

*(b)*

Fig 5: Tip blunting effect in nanoindentation (a) Load depth and (b) hardness

The load-depth (P-h) relation during the initial portion of the curve is P=kh which with increase in depth changes to $P=kh^{3/2}$ and ultimately to $P=kh^2$. Therefore the slope of the curve during the loading is a combination of three functions of depth

$$\frac{dP}{dh} = k + \frac{3}{2}k\sqrt{h_1} + 2kh_2, \text{ where } h_2 > h_1 \quad (4)$$

For films of higher hardness however, for the $P=kh^m$ occur where m > 2 .i.e slope of the curve should increase further and we have to provide higher loads for penetration of smaller depths compared to lower indentation depths. The substrate influence however comes into existence during the unloading. However after a certain depth. The slope of the unloading curve is the measure of stiffness and is used to calculate the elastic modulus.

Dislocation are generated below the contact are and propagate by shear stress at $45^o$ to the loading axis. A harder susbtrate may cause barrier to the dislocation motion casuing their confiment as a result and the higher pile up may occur compared to bulk materials (Fig 6).

Work hardening on the other hand also leads to reducing the dislocation mobility causing the dislocations to move to the surface resulting in higher pile up. If there no or very less work hardeing or there is comparatively not so hard substarte, there is no barrier to the dislocation motion resulting in sink-in effect.

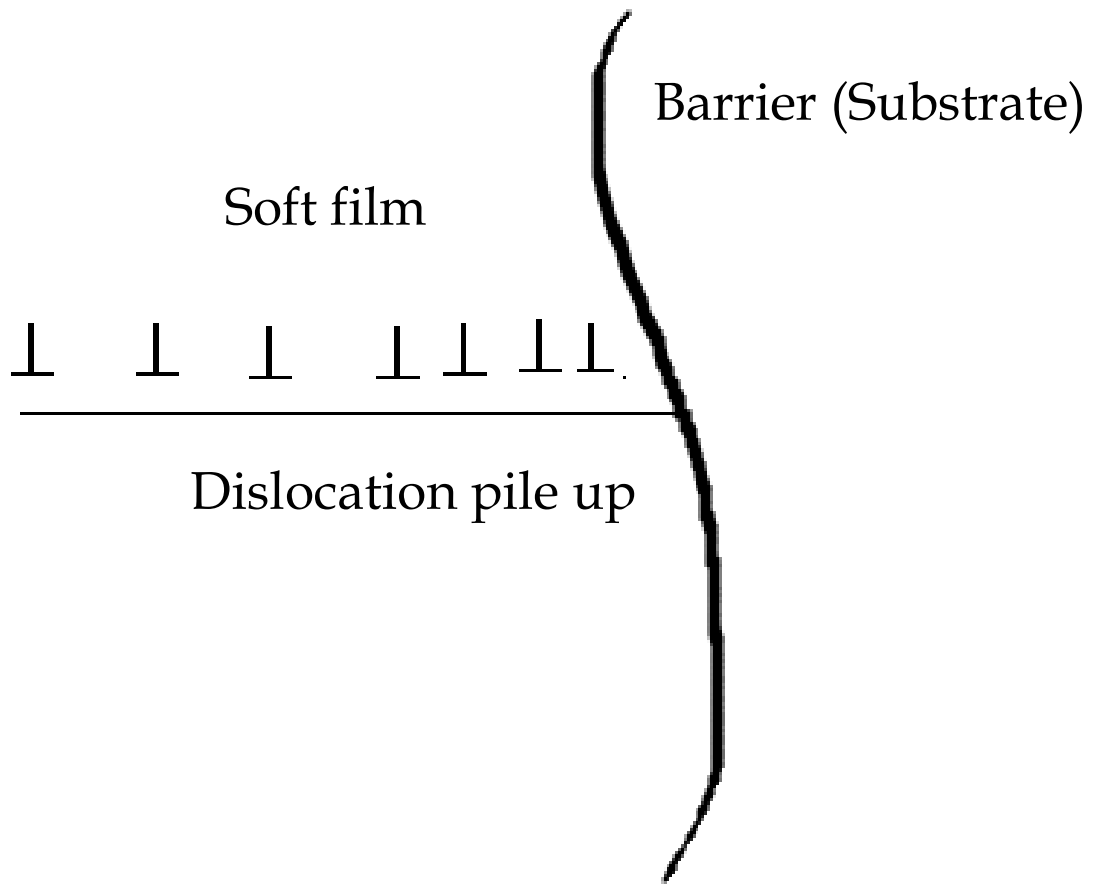

Fig 6 : Dislocation pile up at fil-substrate interface

# 4 Indentation fracture toughness

A nanoindentation performed on Al coatings is shown in fig 7. Pile up is observed surrounding the indentation. Radial cracks were also observed due to the underlying hard silicon substrate. Cracks arise during indentation bulding up of residual stress and begins from flaws which are either preexisting or induced by indentation process itself.

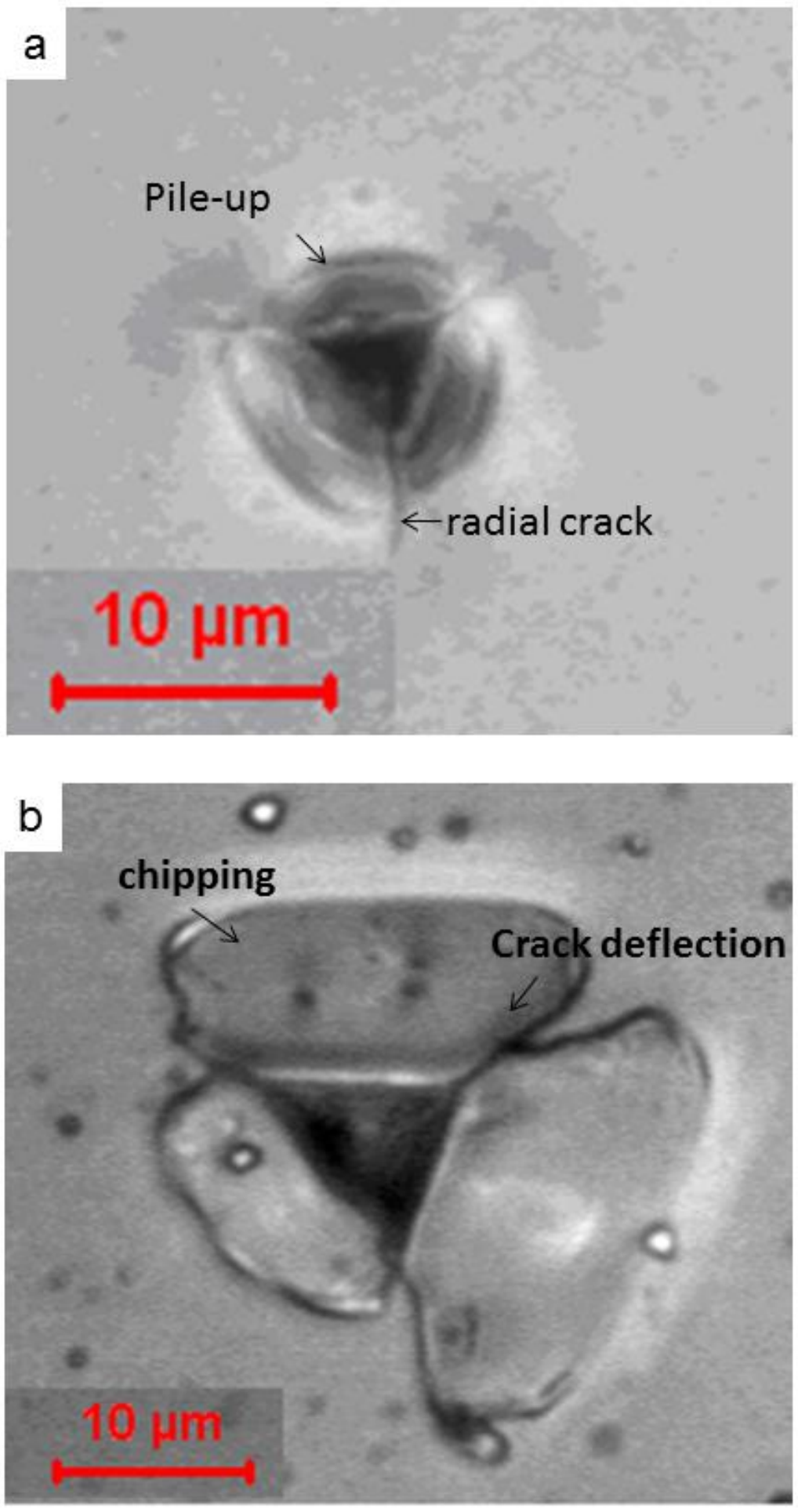

Fig 7 : Nanoindentation of (a) ductile Aluminum coatings (b) hard SiCN coatings on Silicon substarte.

The area of the pileup region can be determined the formula used for segment area of a circle which is given as $\frac{r^2}{2}\left(\frac{\pi}{180}\theta - sin\theta\right)$ where r is segment length from centre and $\theta$ the segment angle  which happens to be the arc length and the cone angle in this case. The area of one pile up region between two radial cracks comes out to be 3.5 $\mu m^2$. Considering the other two similar pile up region, the total pile up area is 10.5 $\mu m^2$. Inhibition of dislocation motion as the indenter encounters a hard subtrate from a from soft coating during penetration and as well as work hardnening lead to pile-up. So the load applied to the pileup area can be consedered as a new parameter called work hardness.

Bucling was observed for hard SiCN coatings on Silicon substrates. The segments formed were also not symmetrical as in the case of pile ups. It is also an indirect estimation of adhesion as poorer is the adhesion, more will be the buckled area. The load applied divided by the area of the buckeld region again gives the adhesive strength or in other words force required to delaminate unit area of the coating. It is also The assymmetrical areas along the three sides again is indicative of the fact that the adhesion strength is not uniform through out. The more is the area difference, the lower is the uniformity. So to determine the adheivs strength we should take the avaerage area into consideration. The adhesive strength multiplied by  sqaure root of the indentation dimension will provide us with indentation interfacial fracture toughness ($K_{IIC}$).

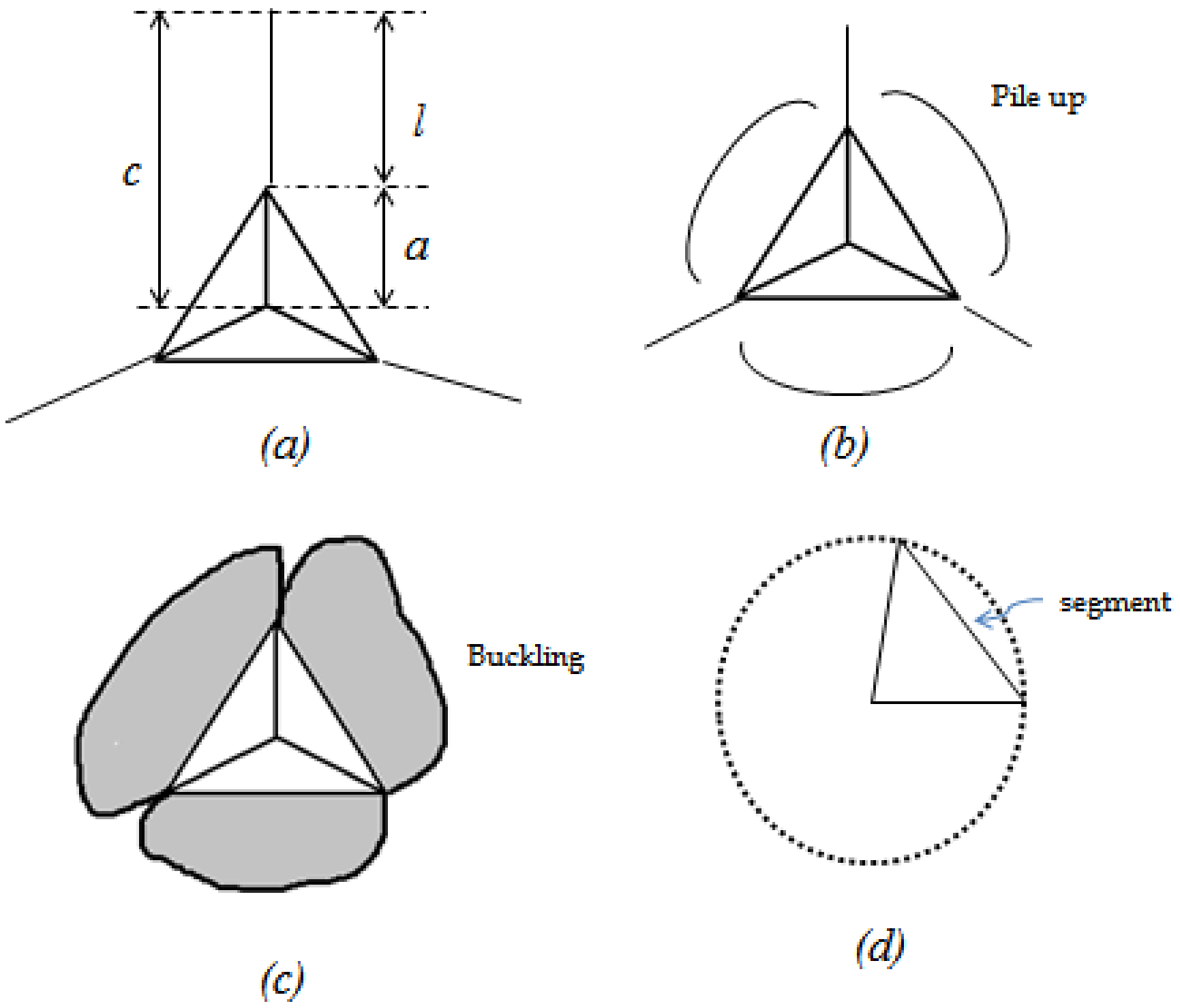

Fig 8: Crack and indentation dimension (b) pileup (c) delamination-buckling around indentation and (d) segment [14]

A higher loads, the substrate effect was much more dominant and chipping was observed. To assess the fracture toughness we use indentation fracture resistance ($K_R$) which depends on the crack morphology and is expressed mathematically based on half penny geometry of palmqvist crack as

$$K_R = \in_R \left(\frac{E}{H}\right)^{1/2} \frac{P}{c^{3/2}} \qquad (5)$$

Where $\in_R$ is a materials constant (~ 0.01), E and H are modulus and Hardness, P is the load applied and 2c is the crack length [15]. The fracture resistance was around 2 MPa √m for a load of 50 mN.

Another equation by Laugier which takes into consideration the crack morphology is given in eqn 6 where $a$ and $l$ are dimensions of imprint and crack as shown in Fig 8 [14, 16].

$$K_R = \kappa \left(\frac{l}{a}\right)^{-1/2} \left(\frac{E}{H}\right)^{1/2} \frac{P}{c^{3/2}} \qquad (6)$$

The amount of pile up occurring surrounding the indentation impression may also paly its role in fracture assessment. The pile up will usually occur if a ductile material is under test or either of the coating and substrate is soft.

# 4 Computational Studies of $K_R$

The indentation fracture resistance ($K_R$) is a function of E/H ratio, indentation load and crack length. A simulation study based on MATLAB programming was done as shown below.

```
e=6:10;
d = 0.5:0.5:5;
P=10:10:100;
c=1:5;
for i=1:10
for j=1:5
Kr(i,j)=0.01*e(j)^1/2*P(i)*c(j)^(-1.5);
Kr2(i,j)=0.01*d(i)^(-1/2)*e(j)^1/2*P(i)*c(j)^(-1.5);
end
end
plot(d,Kr)
plot(P,Kr)
```

```
plot(c,Kr)
plot(e,Kr)
```

The variations of $K_R$ with the different parameters are shown below (Fig 9-15). It can be observed that the variation in Kr is more pronounced at lower values of E/H and crack length. The ratio E/H is an indication of adhesion. Thus coatings with better adhesion will show less variation in indentation fracture toughness.

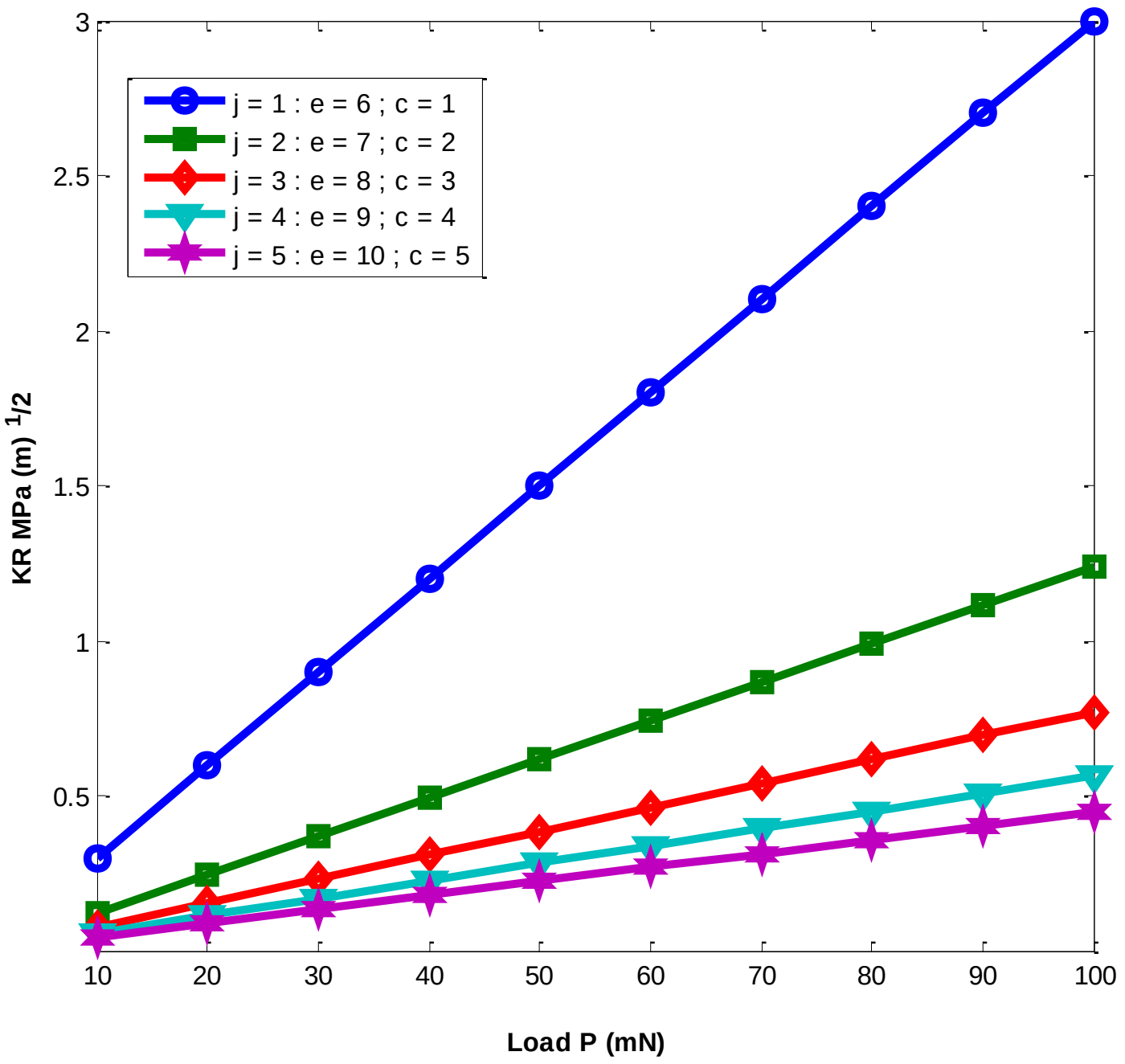

Fig 9 : Variation of Indentation fracture toughens with different loads at different values of E/H ratio and crack length

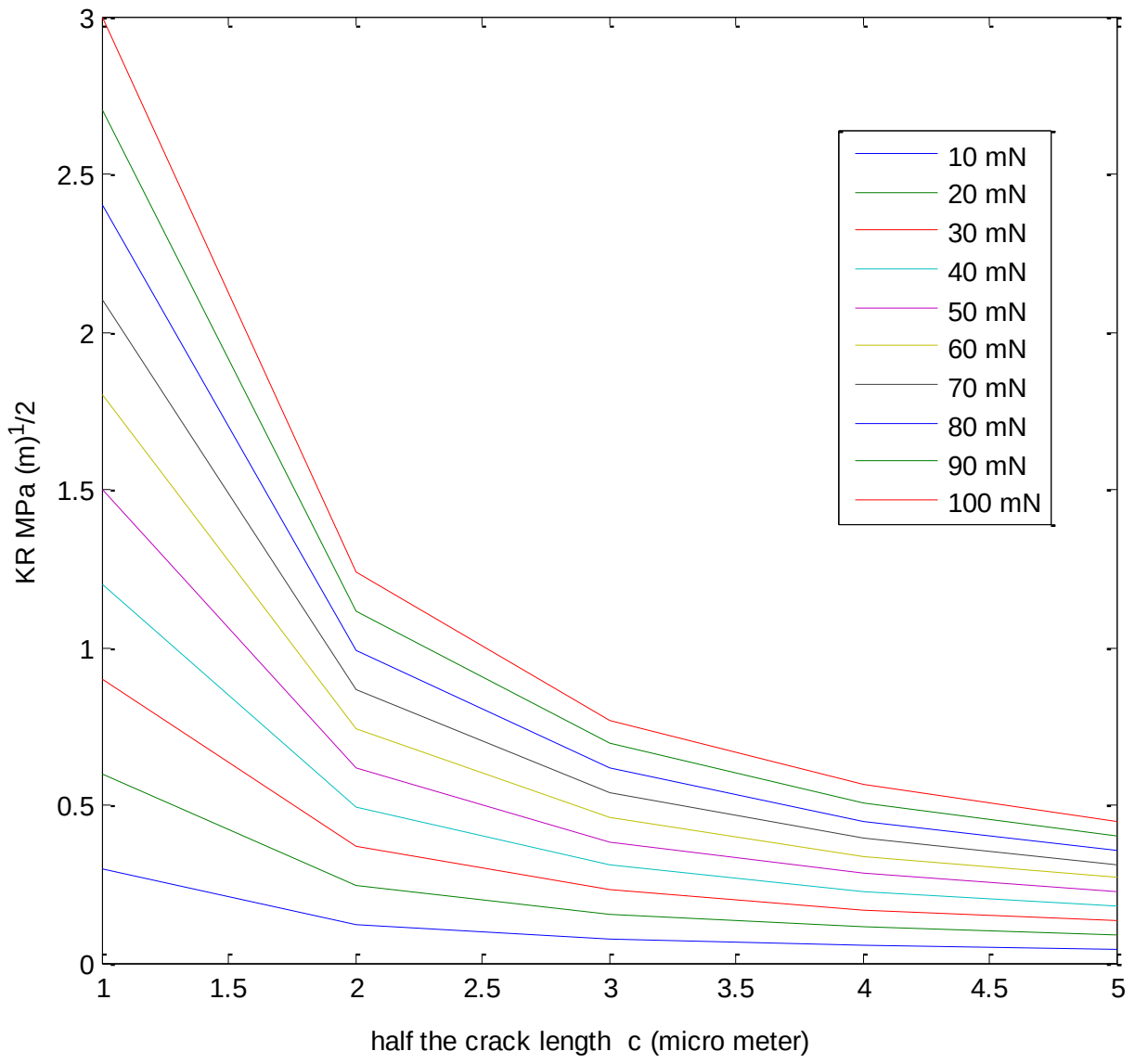

Fig 10: Variation of Indentation fracture toughens (Antis $K_R$) with crack length at different loads

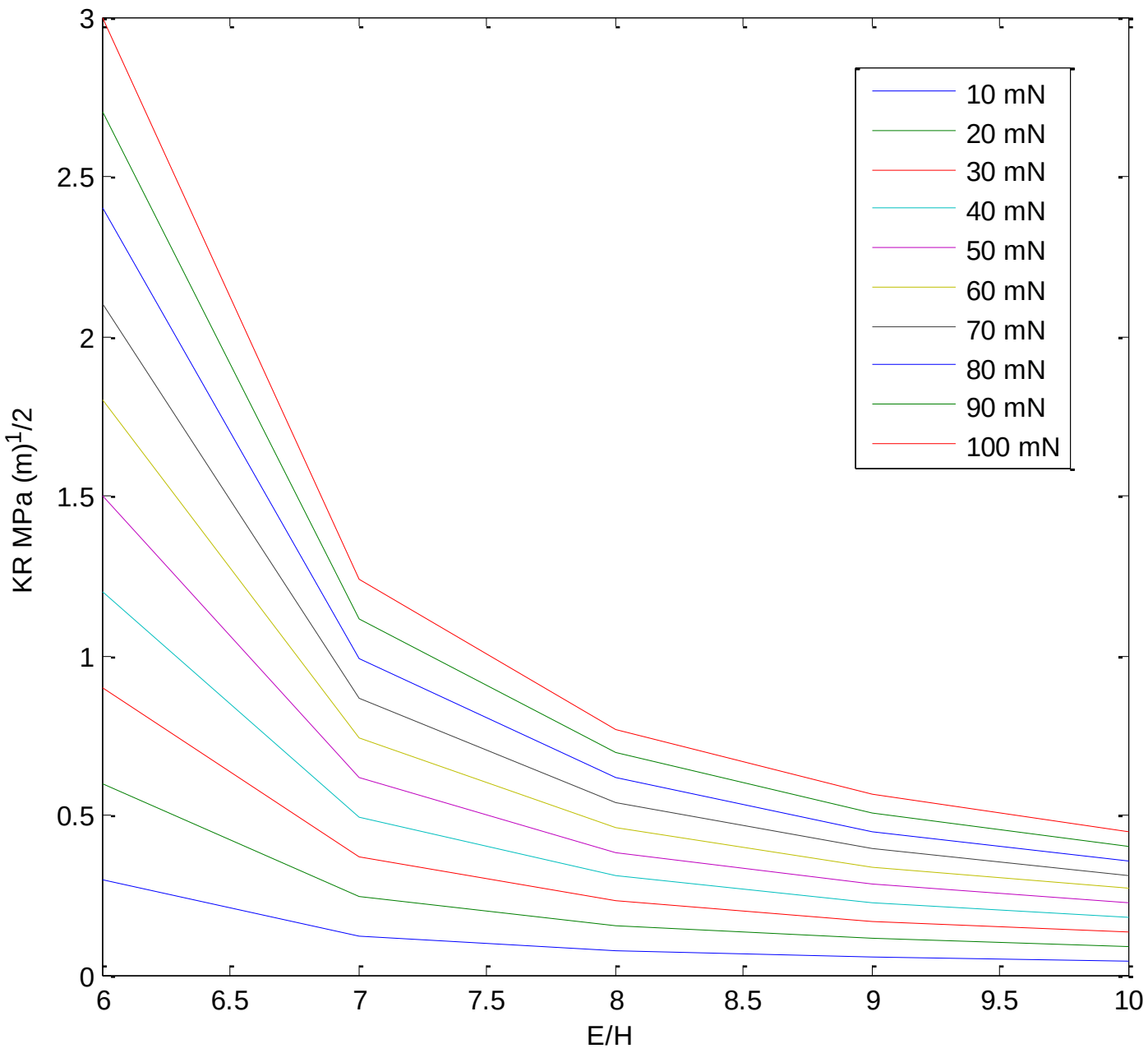

Fig 11 : Indentation fracture toughness (Antis $K_R$) variation with E/H at different loads

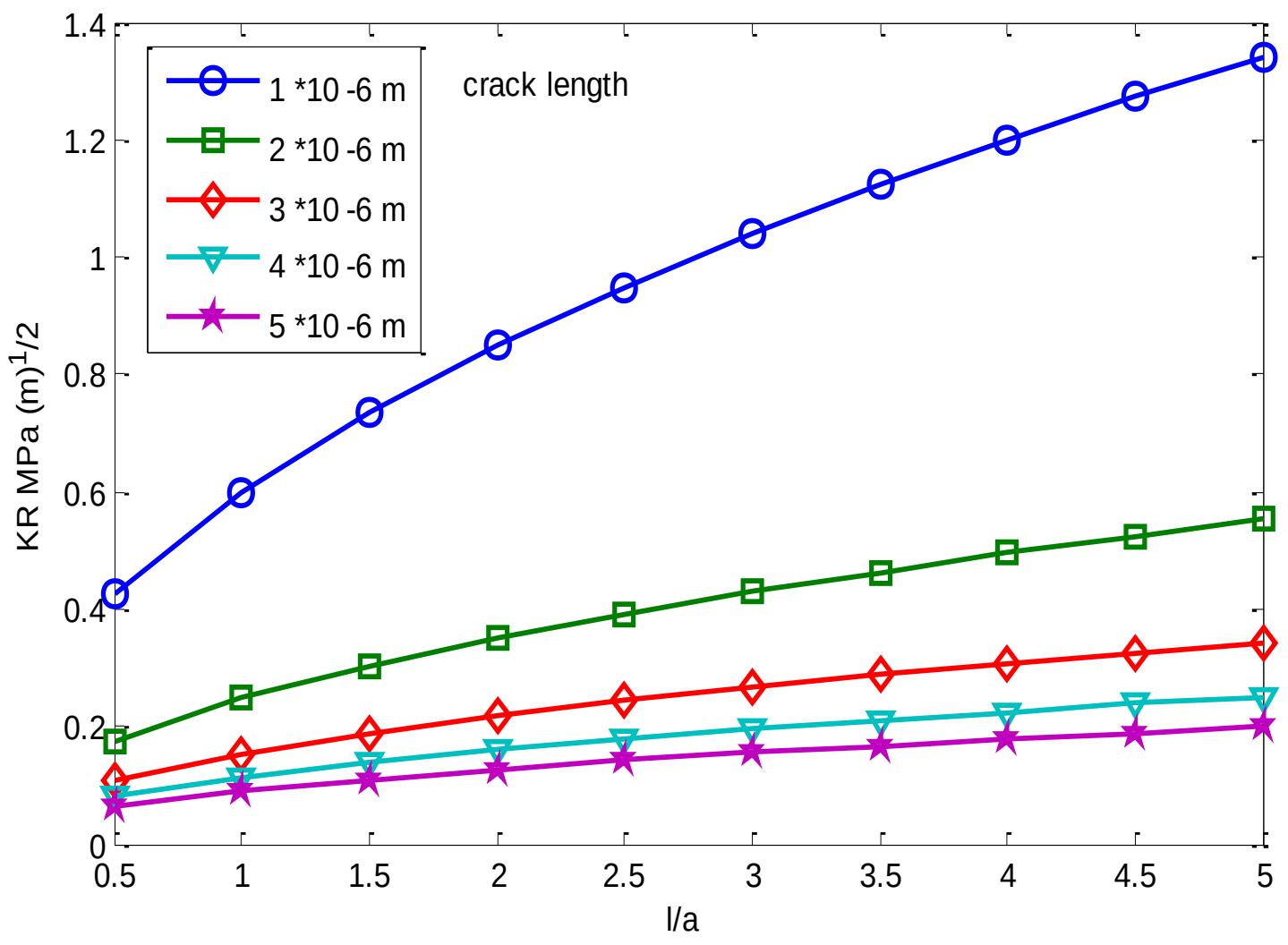

Fig 12 : Indentation fracture toughness, Laugier $K_{R2}$ vs. l/a

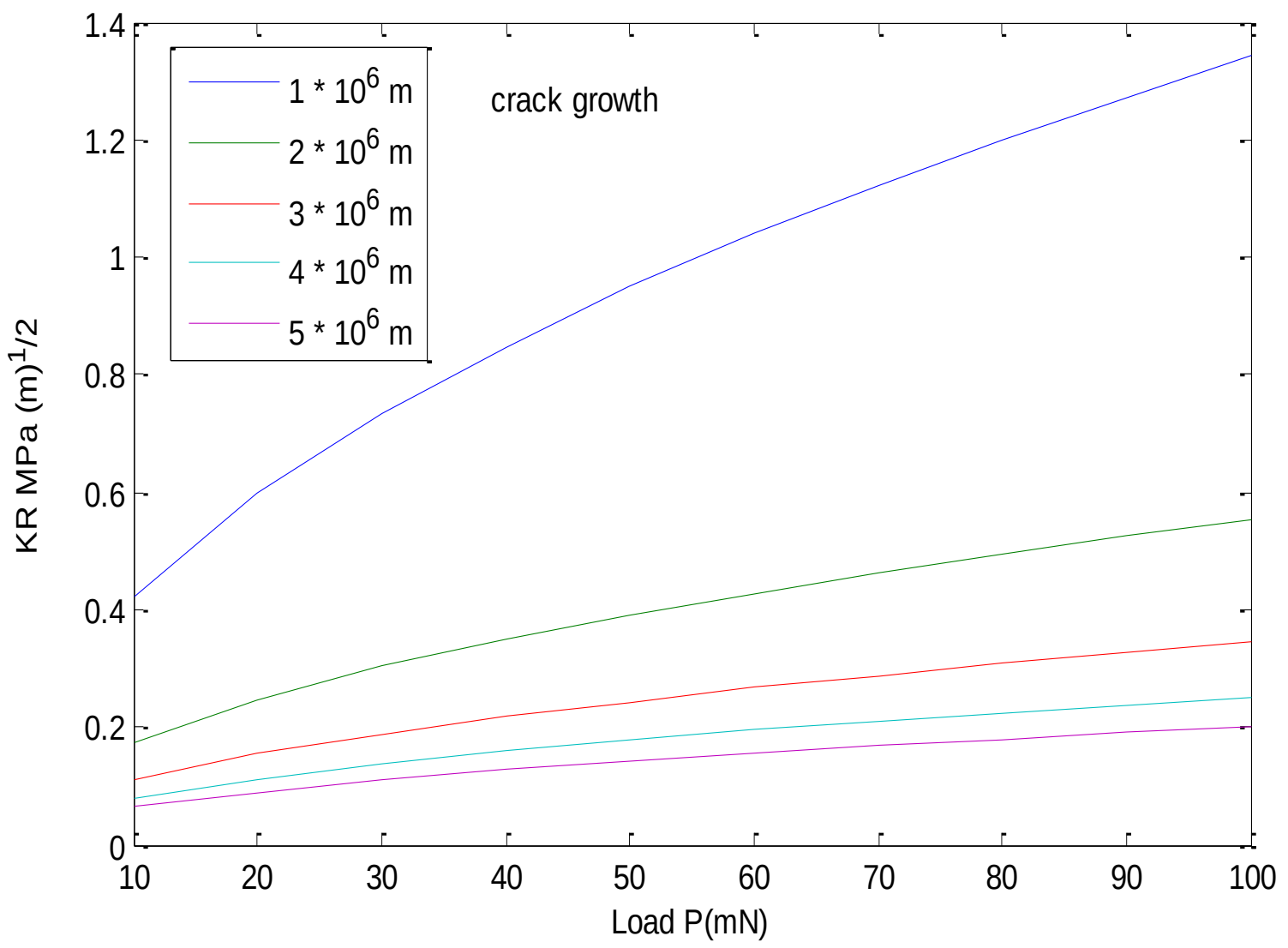

Fig 13: Indentation fracture toughness, Laugier $K_{R2}$ vs. Load P (mN)

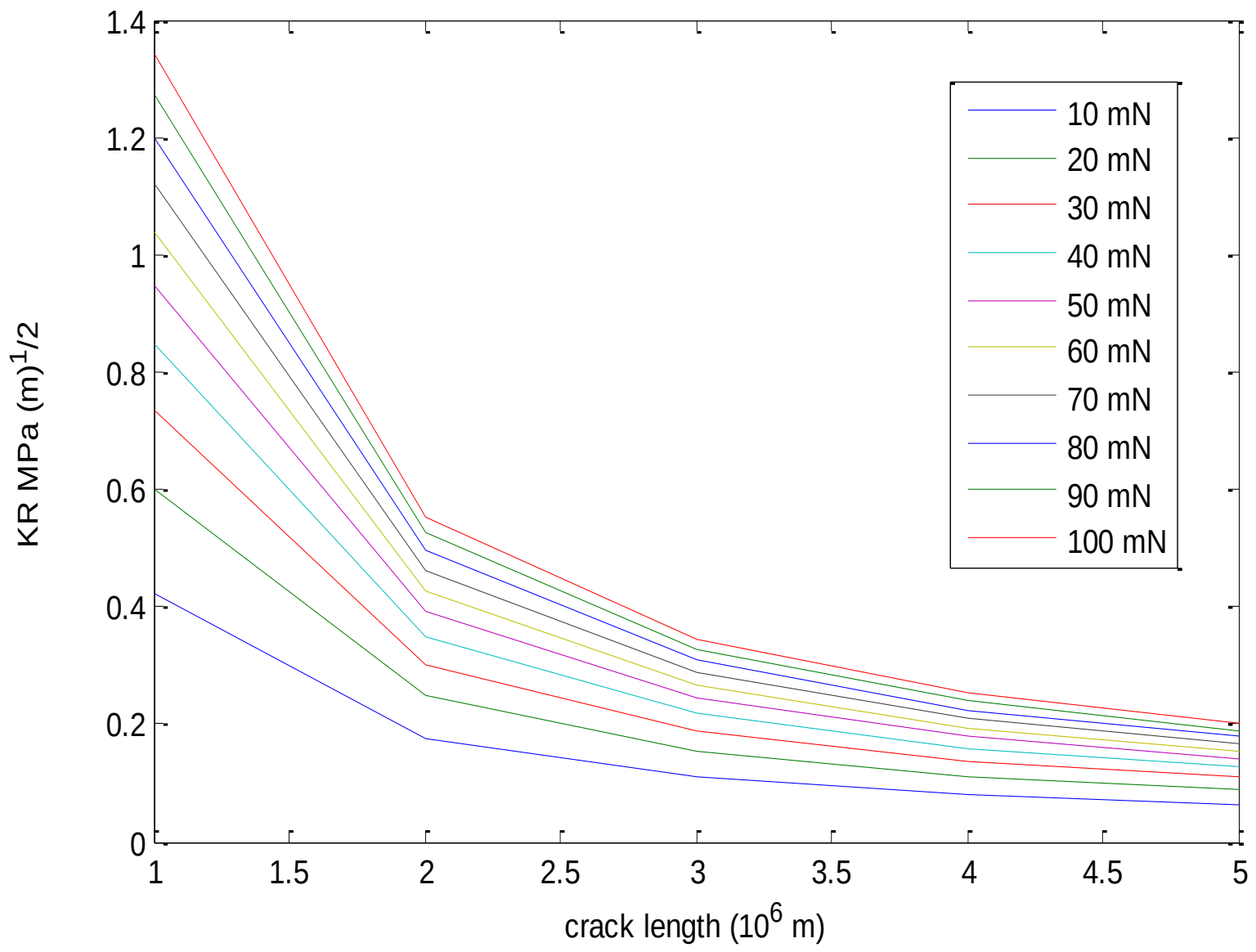

Fig 14: Laugier $K_{R2}$ vs. crack length (µm)

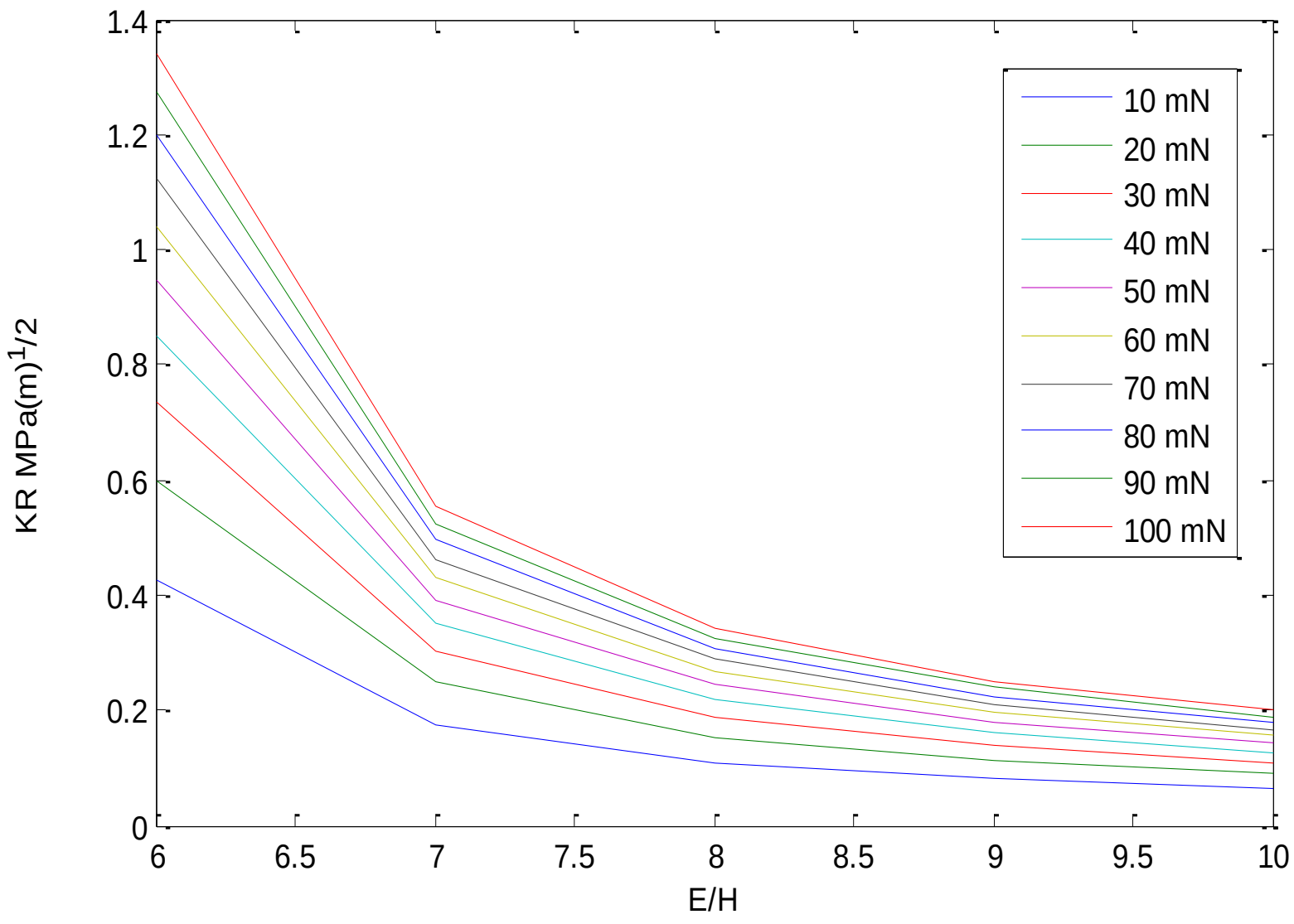

Fig 15 : Indentation fracture toughness, Laugier $K_{R2}$ vs. E/H

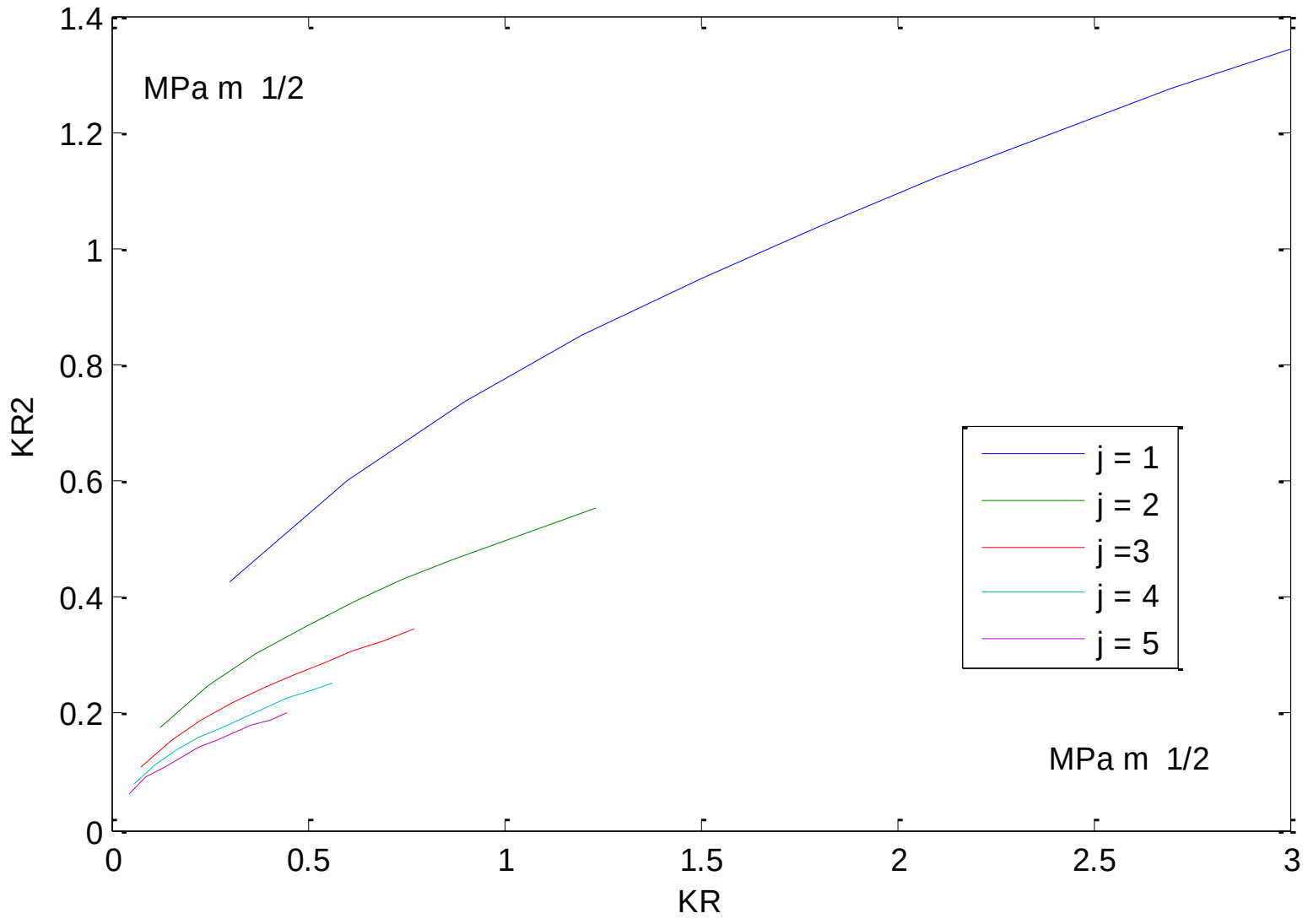

Fig 16: Antis ($K_R$) vs Laugier ($K_{R2}$) indentation fracture toughness

A comparison of indentation fracture toughness developed by Antis ($K_R$) and Laugier ($K_{R2}$) is shown in Fig 16. Firstly the variation is not linear and we are also getting a much lower value of $K_{R2}$ corresponding to $K_R$. The plot shows that for lower values of crack length and E/H ratio, the variation in $K_R$ and $K_{R2}$ is much more pronounced and over a wider range. The results are different from the one published recently [14].

# 5 Computational Studies – substrate effect in indentations at micro level

For hard coatings on comparatively softer substrates, the Vickers hardness usualy is found to decrease with decrease in thickness due to increased substrate effect whereas for coatings of same thickness, the hardness value is found to decrease with increased indentation load again due to substrate effect. The hardness attains saturation in both the cases .The same trend has also been found in the case of hard nanocomposite SiCN coatings.Details regarding hard SiCN thin films can be found in ref [17] and references within.

Hardness expressed as VHN (Vicker's Hardness Number) can be converted to GPa using (100 VHN=1GPa).The hardness profile was fitted with quadratic fit for glass and 304 SS substrates and cubic fit for Si substrates (Fig 17, 18). The fitting relations are given in eqn (7) and (8)

$$H=P_1d^4 + P_2d^3 + P_3 d^2 + P_4d + P_5 \text{ (4}^{th}\text{ order fit)} \qquad (7)$$

$$H=C_1d^3+ C_2d^2 + C_3d + C_5 \text{ (Cubic fit)} \qquad (8)$$

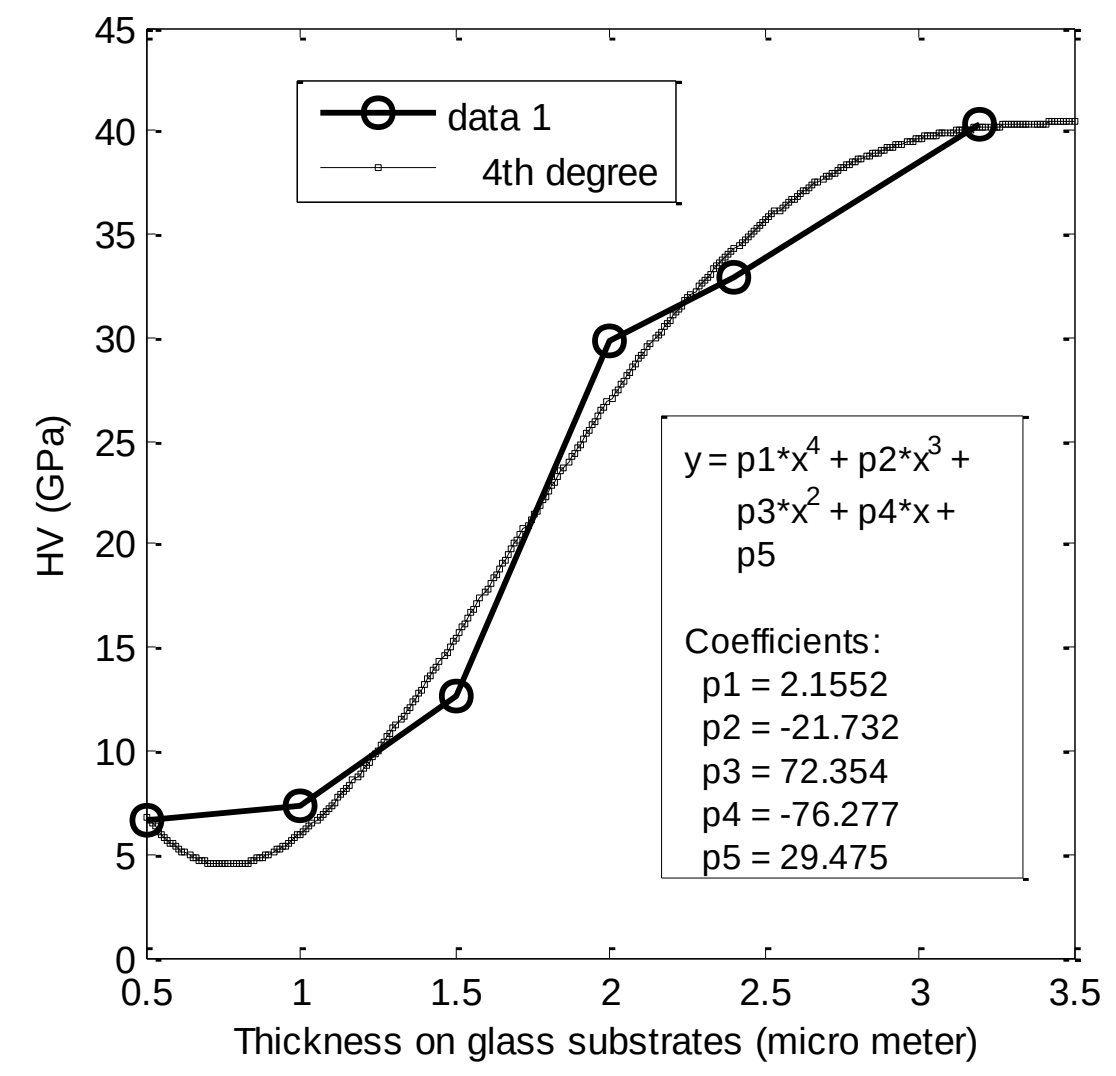

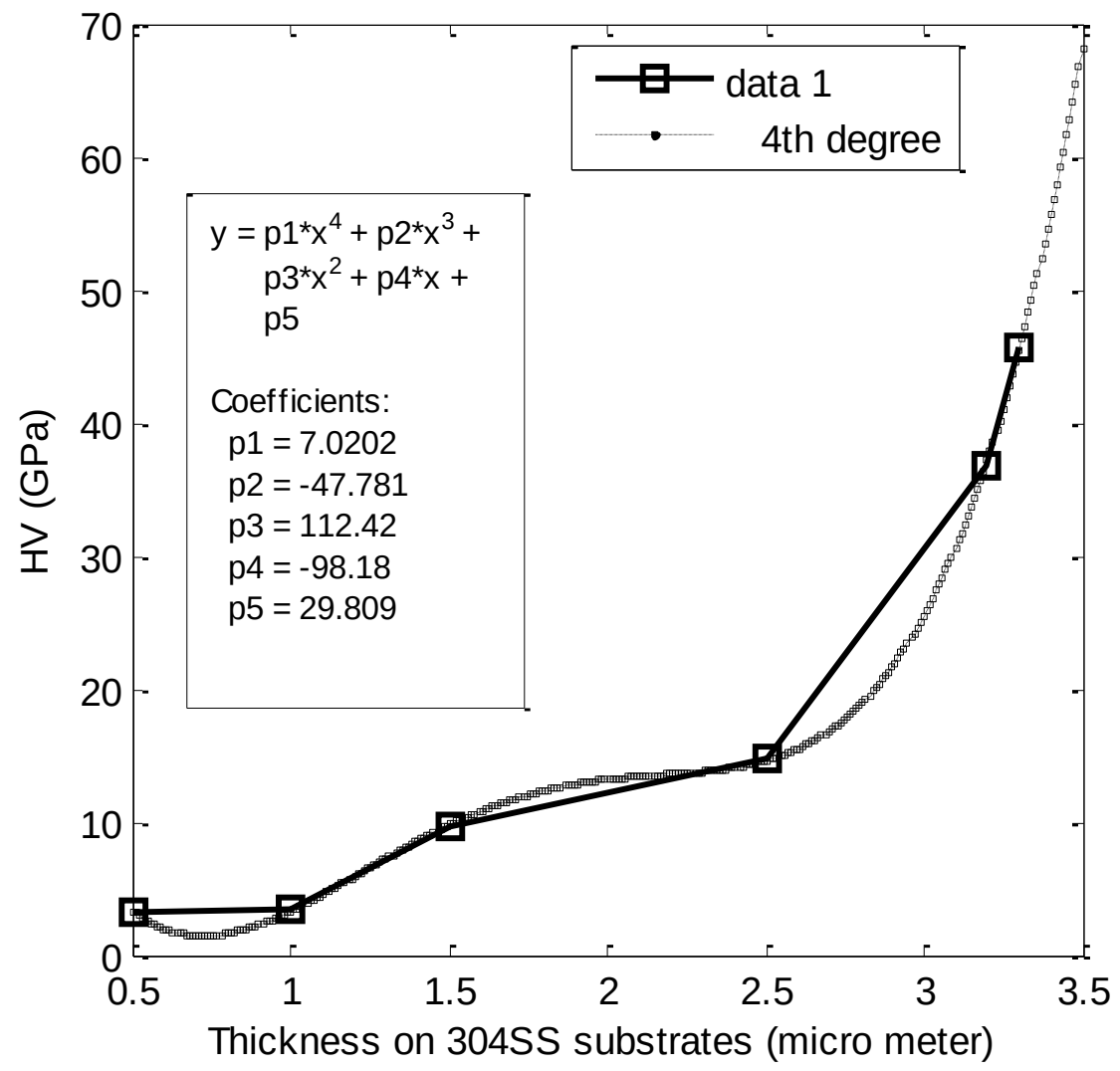

Fig 17: 4th order fit for thin film deposited on (a) glass and (b) 304 SS substrates

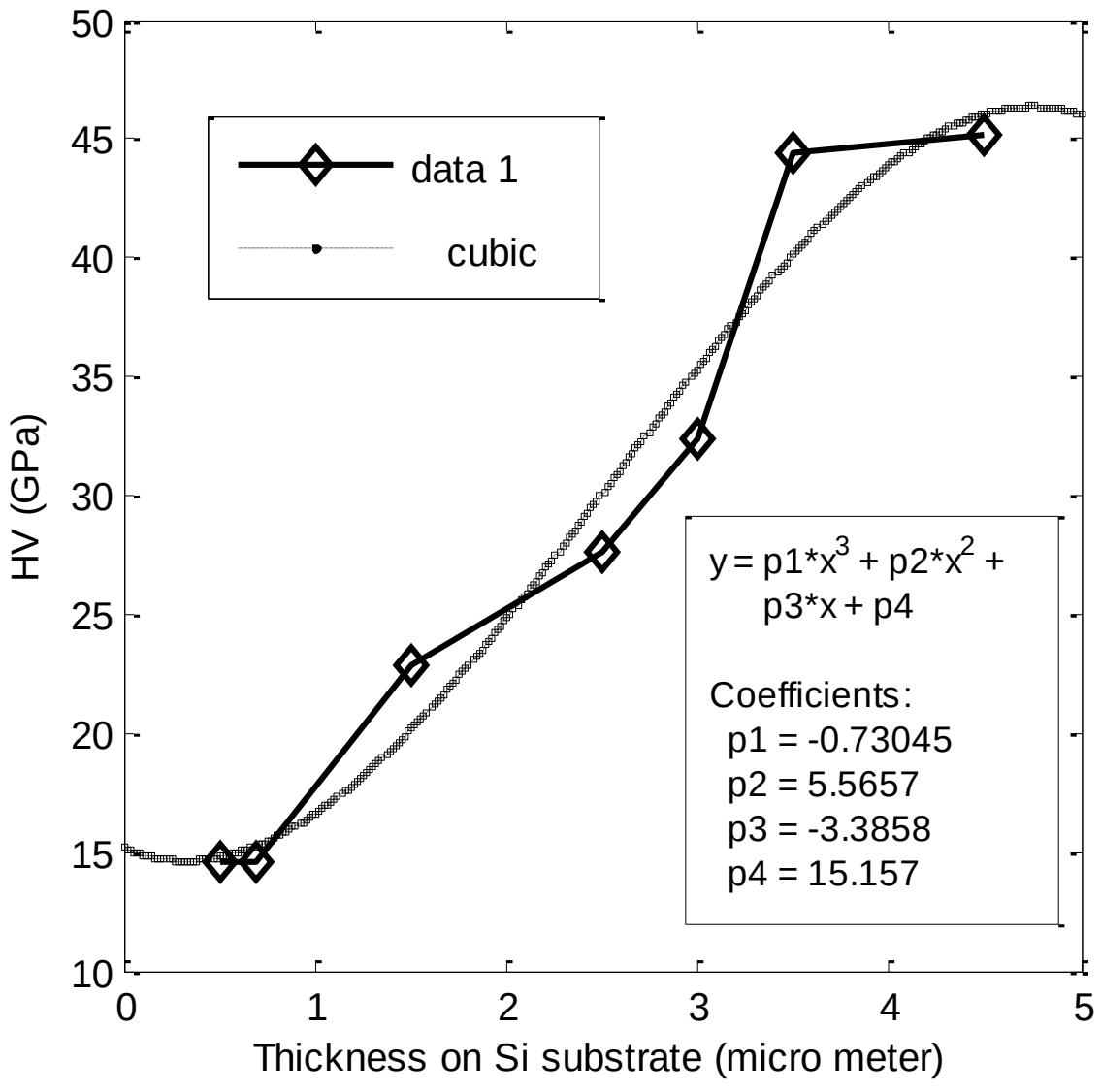

Fig 18: Cubic fit for thin film deposited on Silicon substrates

The parameter P's and C's actually are the ones which quantify the thin film hardness and substrate effect. These values were varied from -100 to 100 with spacing of 0.1 (2001 values) and the thickness was varied form 100 nm to 10μm with an increment of 50 nm (199 values) to find the different hardness vs. thickness profiles. The $P_1$ values were varied from -1.0 to 3.0 and the *H (d)* values were determined and shown in fig 19. The Hardness values are hypothetical and therefore only their change may be considered.As an indenter penetrates a sample, the contact geometry and its corresponding area changes. The hardness was found to remain constant with thickness for a $P_1$ value of 2 approximately (Fig 20). Hardness showed an increase with increase in $P_1$ and decrease with lowering of $P_1$ values. Hardness as we know is the restriction imposed by the material towards movement of dislocation. The increase in hardness with thickness is due to reduced substrate effect. However a decrease with thickness can be explained as poor adhesion due to increased residual stress. Compressive stress is beneficial for film adhesion whereas tensile stress is detrimental. The variation of Hardness with $P_1$ for different film thickness is given in Fig 20. It can be observed that for thickness less than 2μm there is lower evidence of stress variation as the hardness is almost constant. However a higher value of $P_1$ leads to higher stress variation in the material leading to instability.

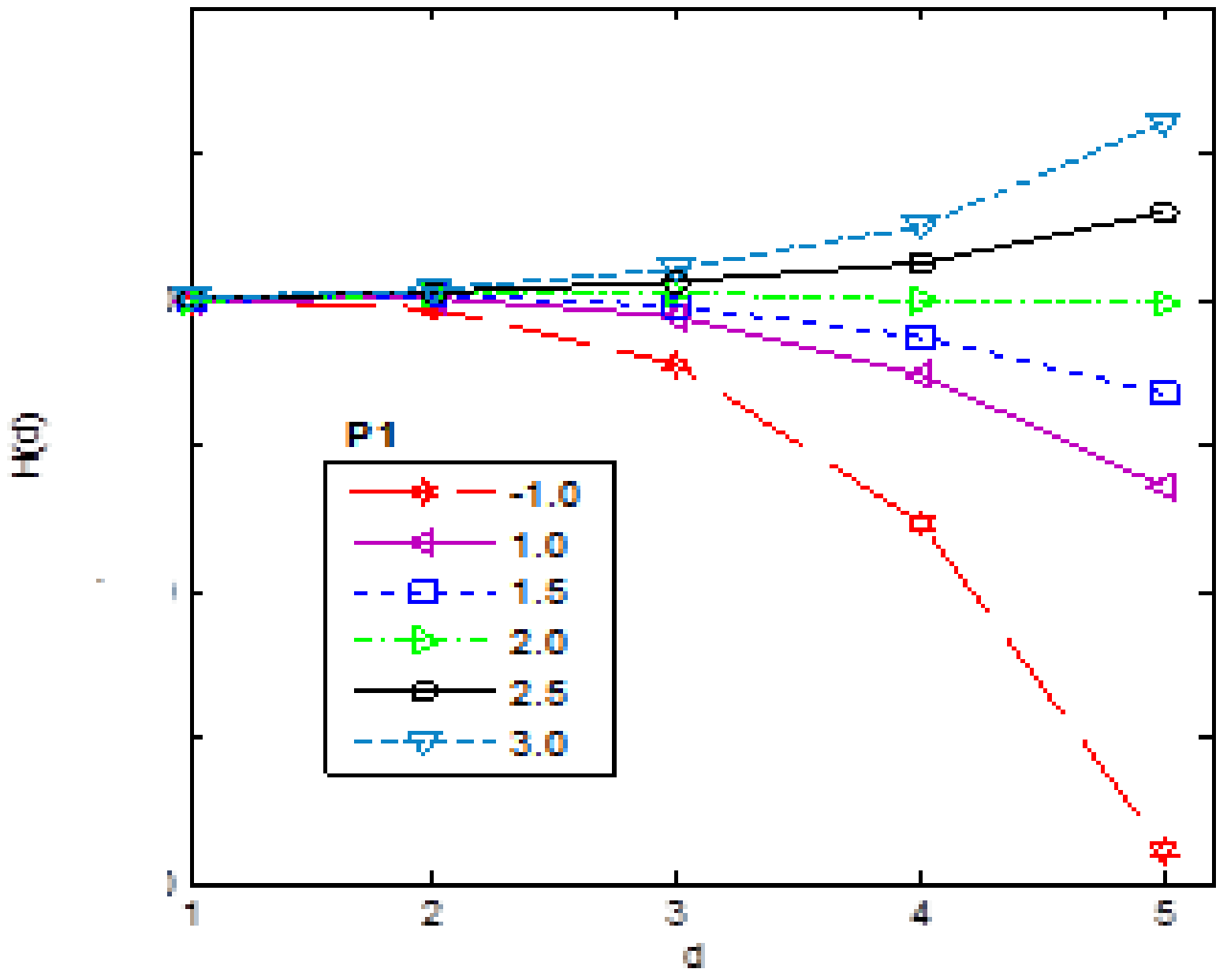

Fig 19 : Variation of Hardness with thickness for

different $P_1$ values

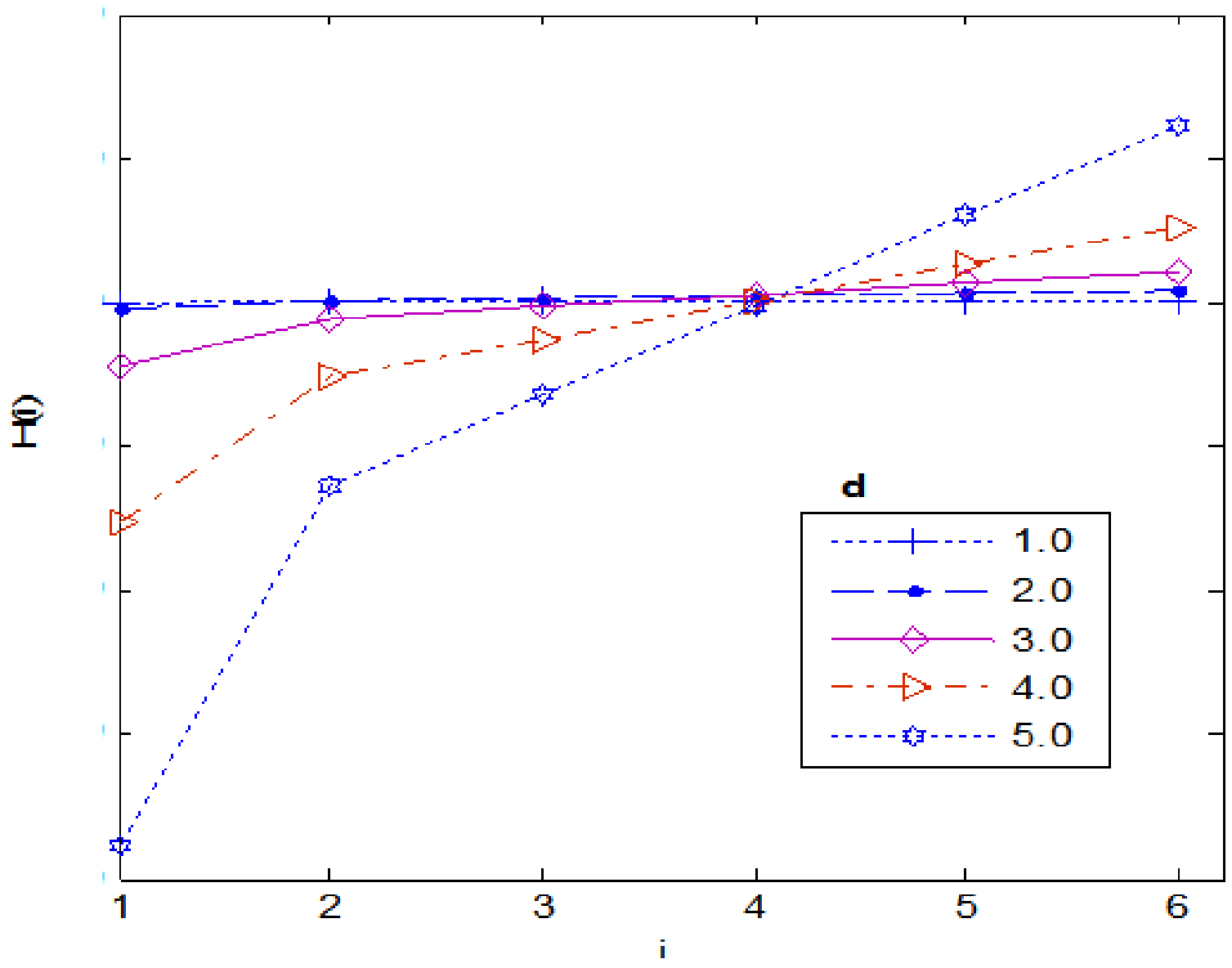

Fig 20 : Variation of hardness with $P_1$ values for different thickness

The variation of hardness with $P_2$, $P_3$, $P_4$ and $P_5$ for different thickness (1- 5 μm) are shown in Fig 20- 24. We can see that there is a decrease in hardness with increase in $P_2$ and the rate of decrease is higher for higher thickness. Interestingly a reverse effect of increase is observed in hardness for $P_3$ and also the rate is not monotonically increasing or decreasing with thickness and shows highest rate of increase for 4 μm.

The difference in the rate of decrease is much lower for $P_4$ and almost constant for P5. We can see that the variation in hardness with thickness was maximum for $P_3$ and least for $P_5$ (Fig 25)

We can associate $P_2$ as well as $P_4$ with tip sharpness as sharper is the tip the lesser will be the hardness. The effect of pile up can also be associated with $P_2$ as higher is the pile up lower will be the hardness.

$P_3$ on the other hand can be associated with tip bluntness, work hardening as well as sink in effect. Consequently the harness shows an increase with $P_3$. P5 is simply a constant hence shows no change with its increase.

The abrupt change for thickness of 4 µm is an indication of optimized thickness showing the best possible properties which has also matched with experimental results [2].

From fig 25 we can also observe that the most dominating parameter has been $P_3$ in this case as the variations are more vivid (Fig 25 (b)) compared to others. Hence we can also say that for SiCN, the sinking and tip blunting have been significant.

However the tip blunting or sharpening is much more significant for nanoindentation [3] and so in this case we shall stick to the fitting P parameters define the material properties more than indentation methodology. This part is also stored in ref [30].

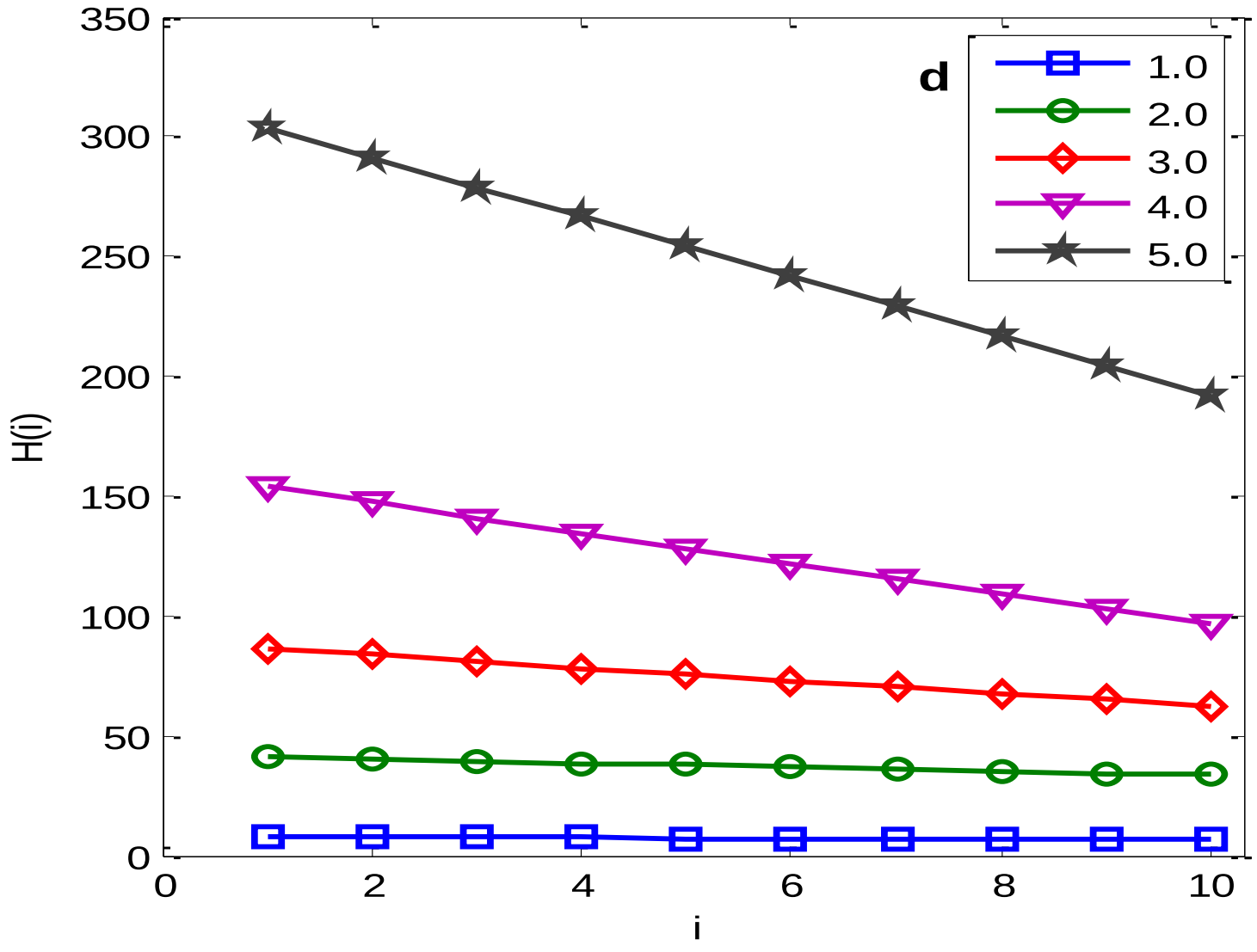

Fig 21 : Variation of hardness with $P_2$ values for different thickness

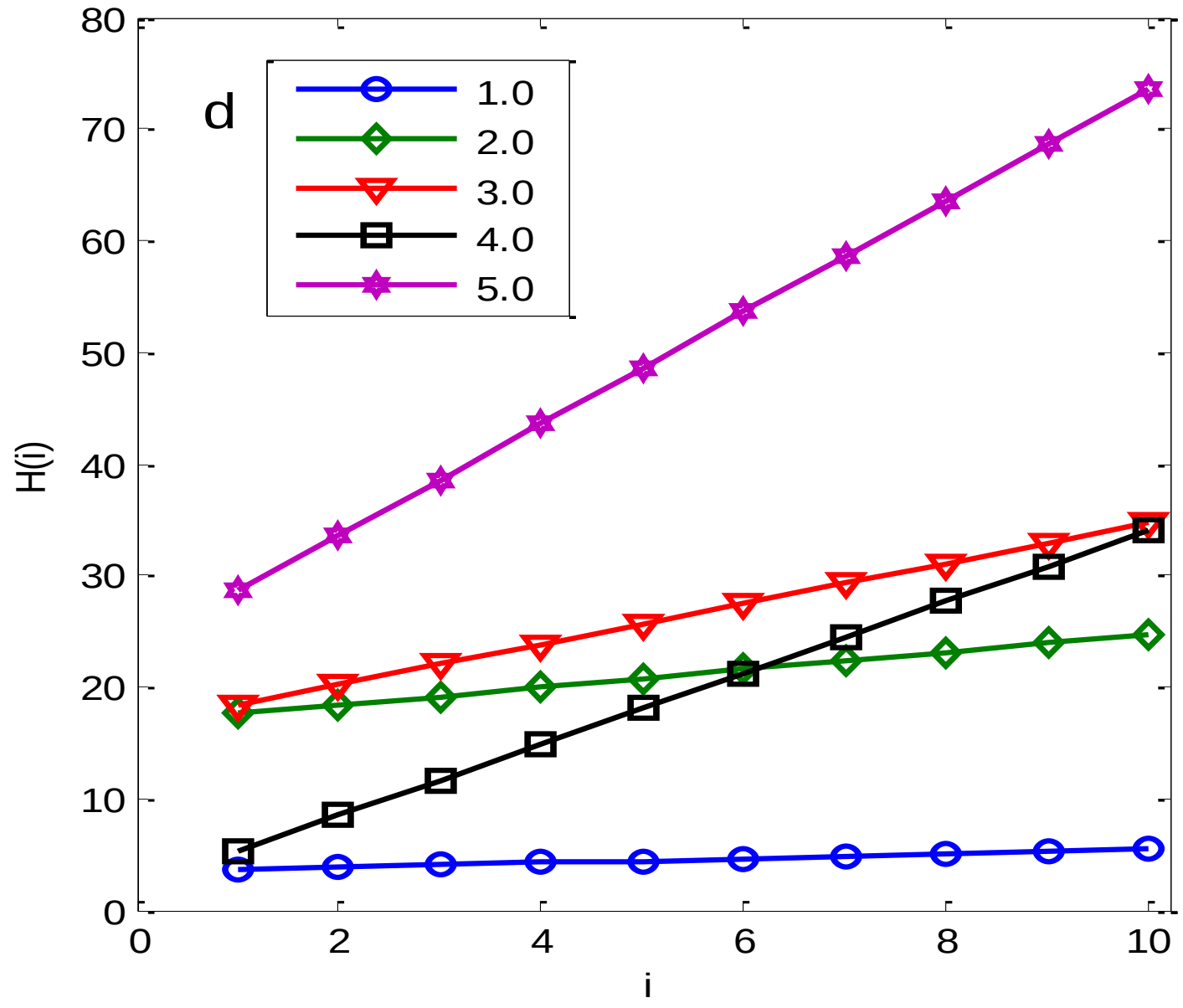

Fig 22 : Variation of hardness with $P_3$ values for different thickness

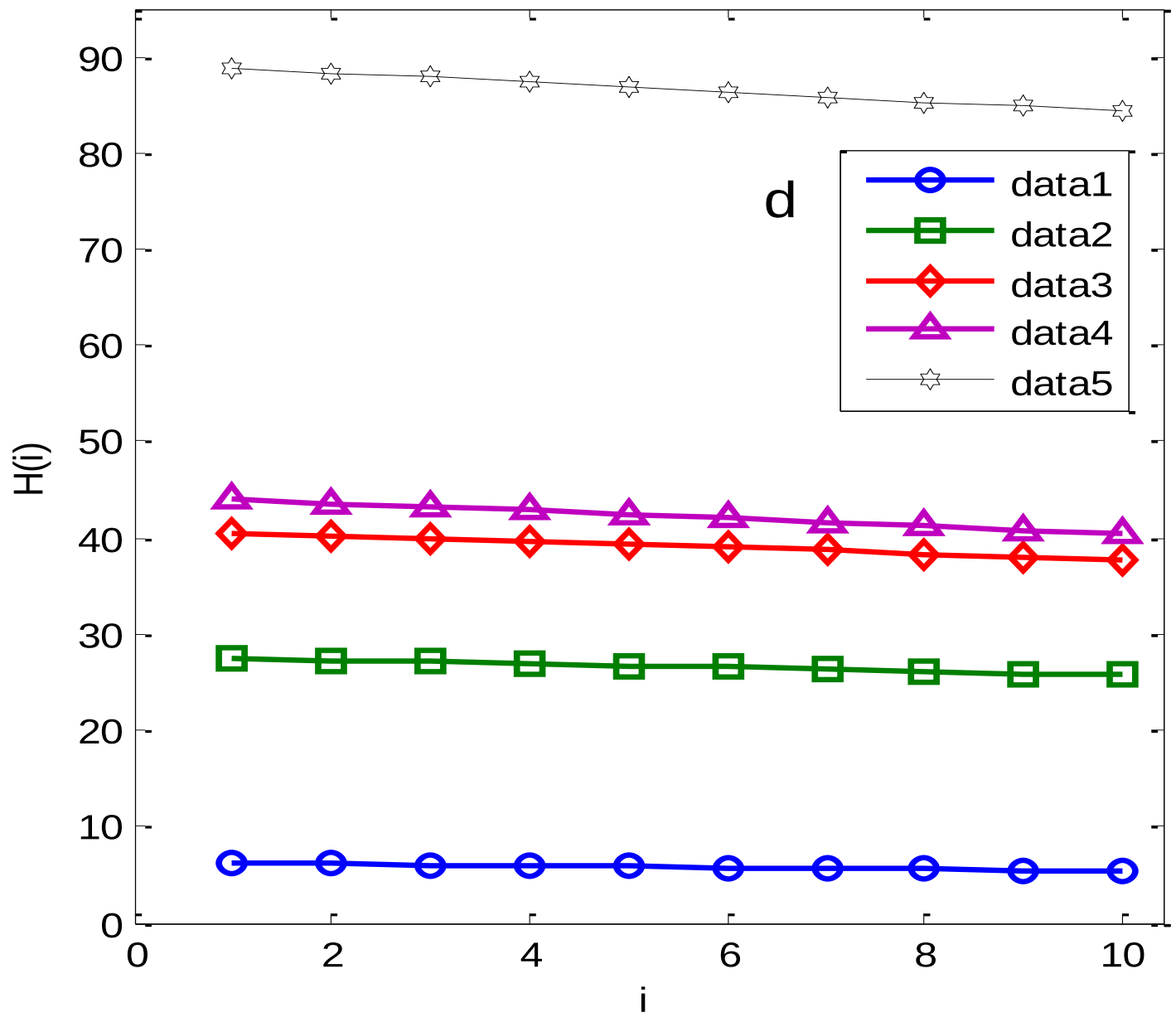

Fig 23 : Variation of hardness with $P_4$ values for different thickness

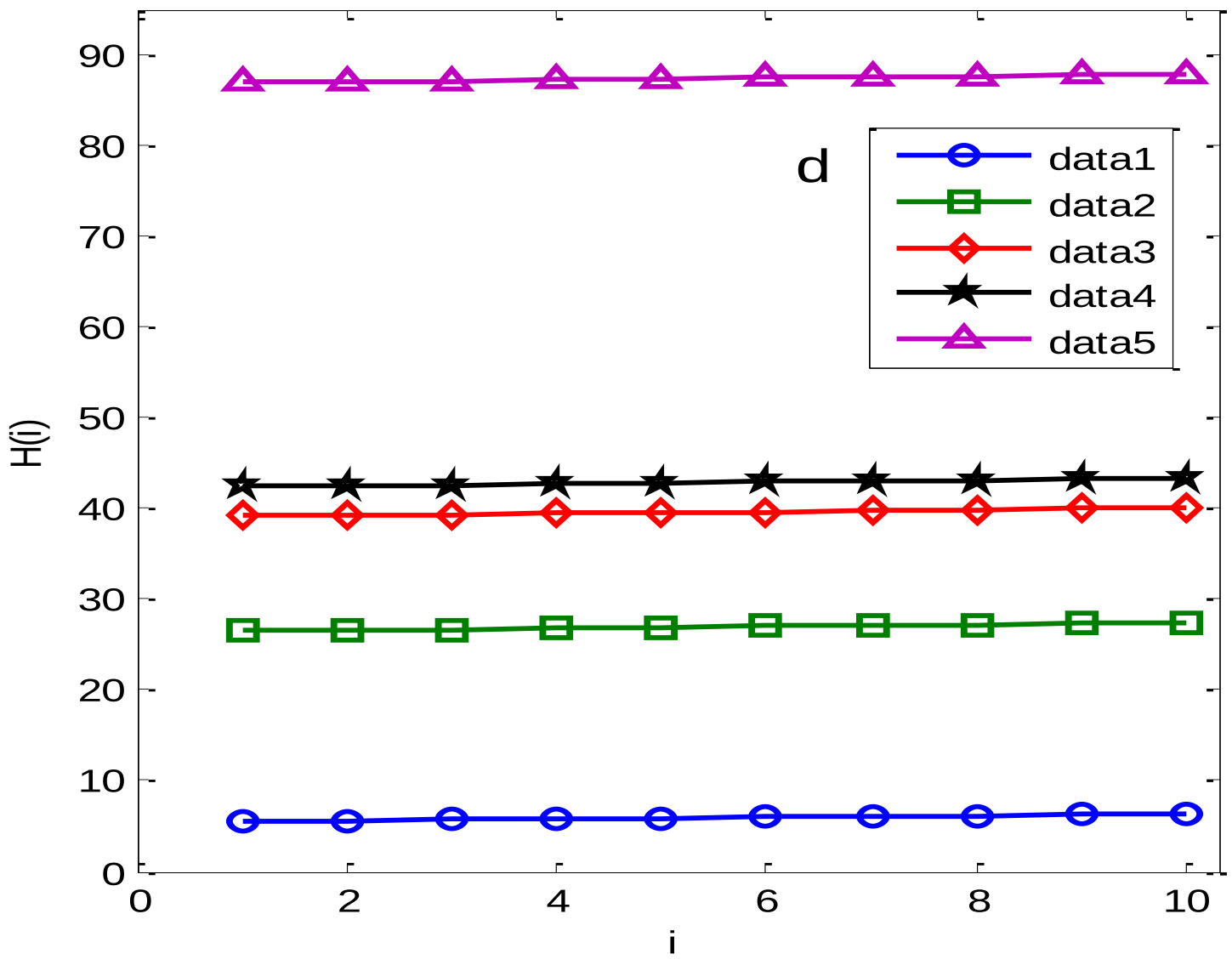

Fig 24 : Variation of hardness with $P_5$ values for different thickness

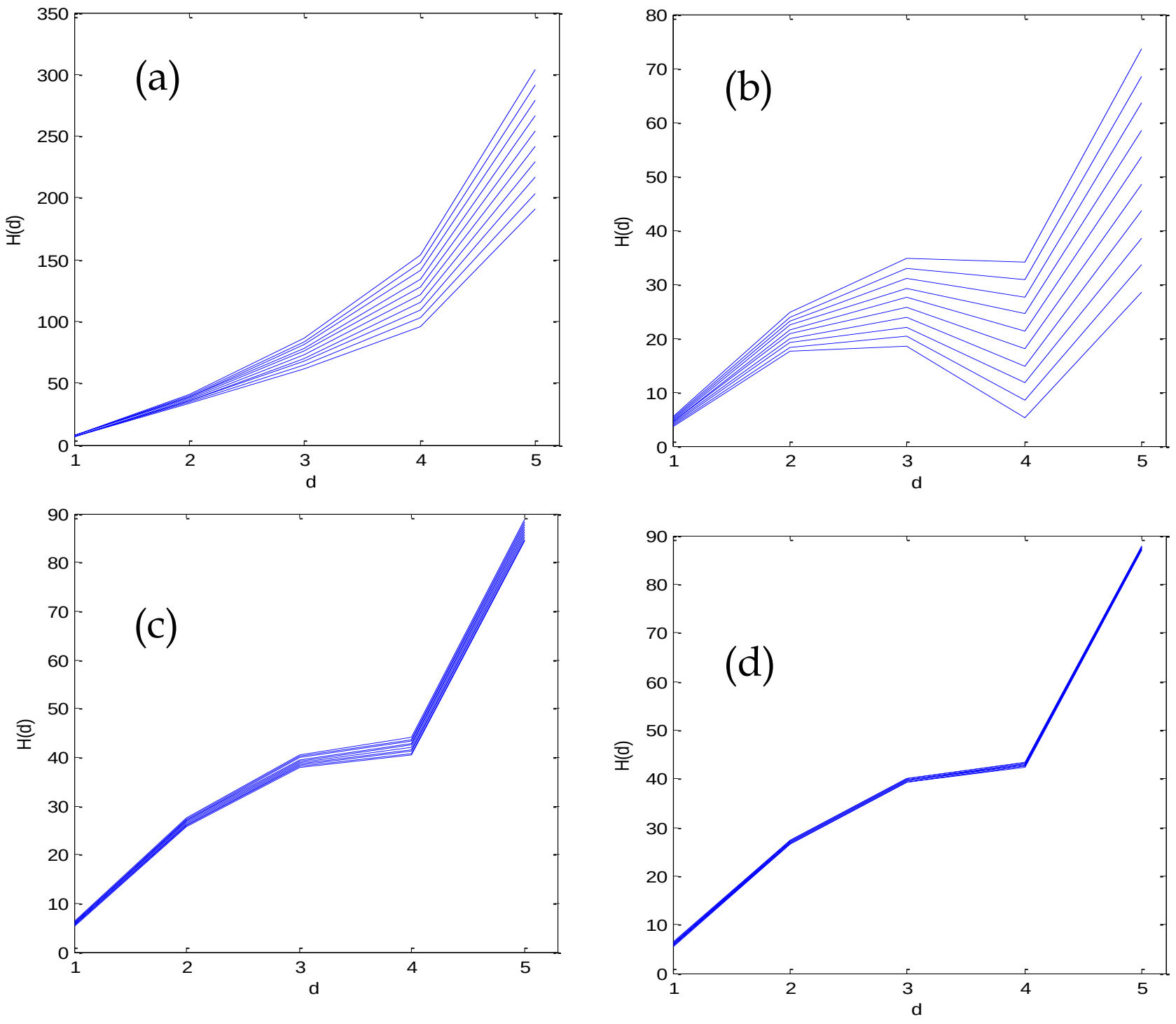

Fig 25 : Variation of Hardness with thickness for different a)$P_2$, b) $P_3$. c) $P_4$.and d) $P_5$ values

# 6 Thin film indentation failure at microscale and delamination area geometry effect in indentations at micro level

Hard thin films are required for the protection of machine parts under wear and abrasion. Although there are a lot of studies on structure and phase formation, the hardness and toughness mechanisms of such thin films as well as their detailed study on fracture behaviour are required. The mechanical properties of thin films be can determined by microindentation and nanoindentation.

Microhardness tests using a Vickers indenter under different loads showed different cracking phenomenon in the thin films deposited on glass substrates (Fig 26(i)). The loads were varied from 50 gf to 1000 gf. A prominent square pyramidal indentation was observed for 50gf accompanied by delamination in the region surrounding the indentation which was due to the flow of shear force. On increasing the load to 100 gf, radial cracks from the corners of the indentation was observed. Picture frame cracking followed by circular shear flow were observed on increasing the load further. A schematic representation of the types of deformation is shown in Fig 26(ii) [17].

The delaminated area surrounding the indentation from 200 gf onwards can be resolved to geometrical shapes or a combination of geometrical shapes. This region corresponds to the enclave of residual compressive pulse which occurs as the plastic zone near the crack tip which has grown in size due to overload since unable to regain its shape after unloading. The delaminated area at 300 gf resembles that of a cardioid [18]. An interesting observation can be made that with increase in load the delaminated area although increasing in size but also meeting the indentation impression ends and forming a cusp. The load divided by the area of delamination can give us the value of compressive pulse which was roughly 4.2 GPa for 300 gf. The area of a cardioid is $3/2\ \pi\ a^2$ where a is the length shown in fig 27(a). The compressive pulse was found to increase with load (Fig 27(b)). A detailed study of crack formation and determination of fracture toughness can be found in ref [18, 19].

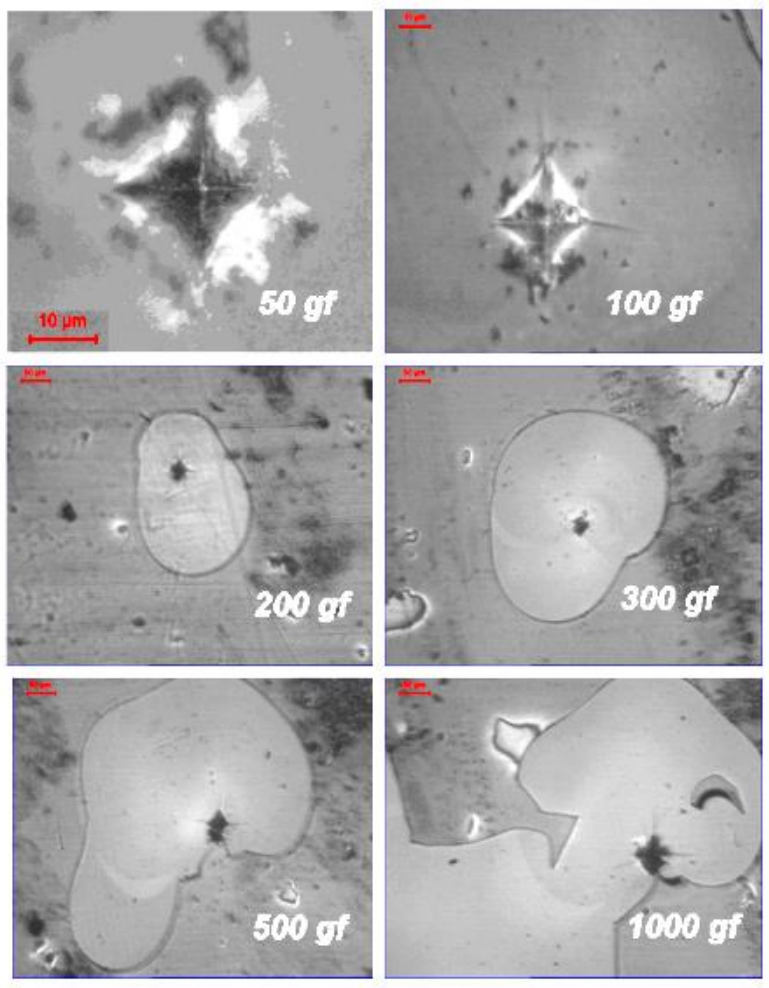

(i)

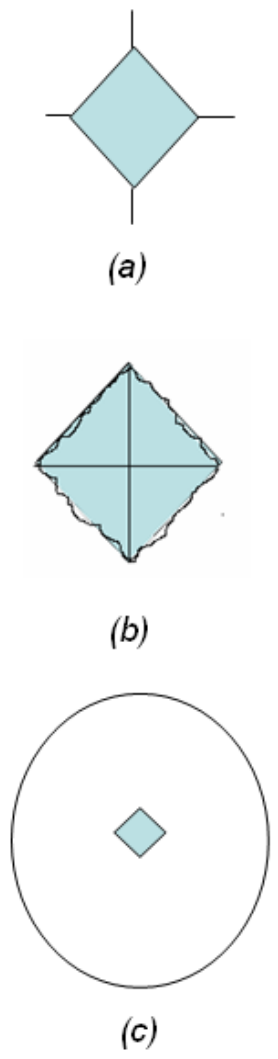

(ii)

Fig 26 (i) Vickers microindentation at different loads on SiCN thin films. (ii) Schematic representation of (a) Radial cracking (b) Picture frame cracking and (c) Circular through thickness cracking [17]

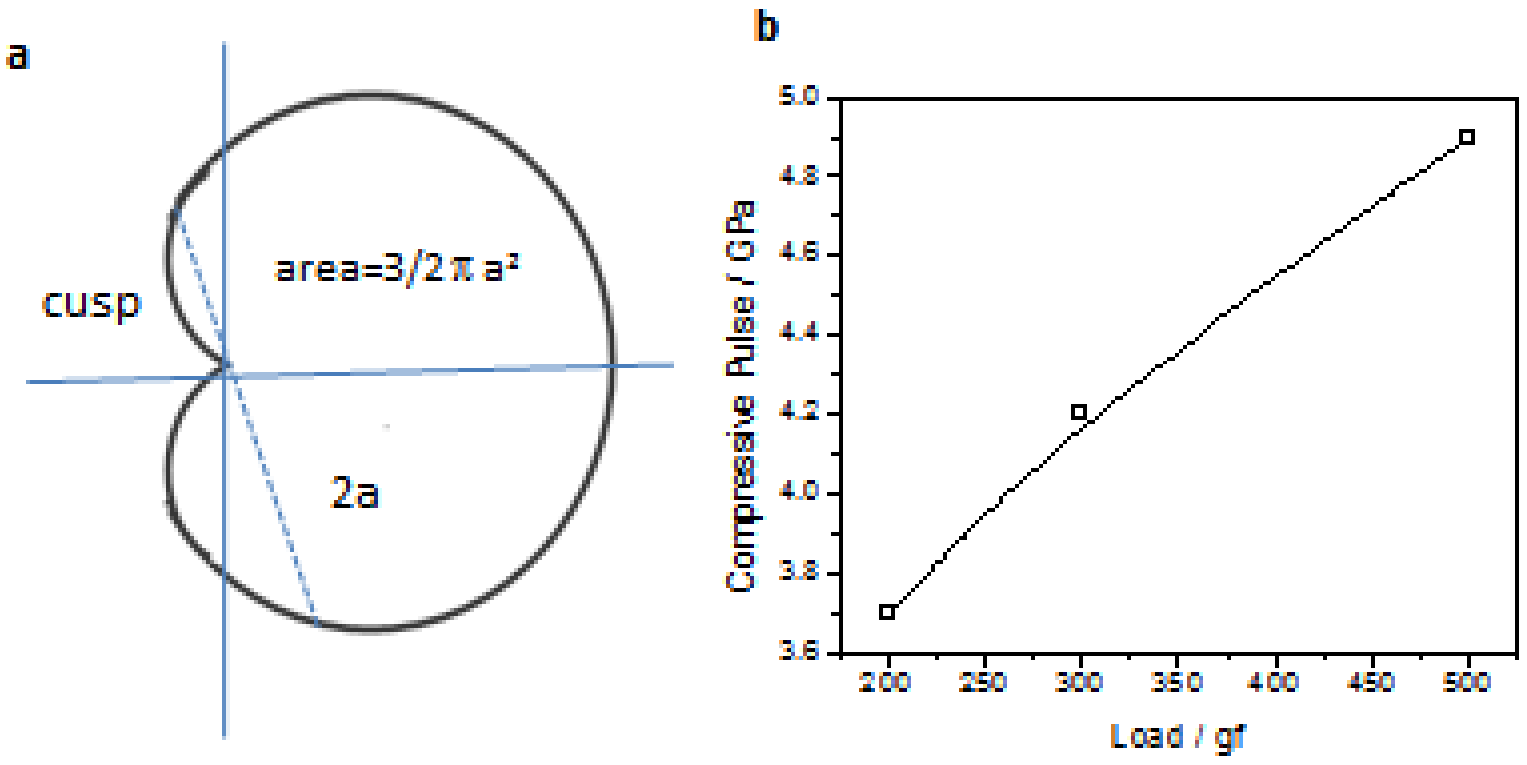

Fig 27 (a) Geometry of a cardioid (b) Compressive pulse variation with indentation load.

Fracture studies using Vicker's indentation was performed on other nanocomposites, single crystals showing cracks, lateral crack chips and delamination. Presence of nanocrystals increased the fracture toughness [20, 21]. Nanoindentation is an effective tool for determination of material properties based on its depth sensing capabilities which are derived in the following manner [22, 23]. During nanoindentation a Berkovich indenter with 70.3° effective cone angle pushed into the material and withdrawn. The indentation load and displacement are recorded. Nanoindentation studies also showed formation of radial cracks from the corners of the indentation impression (Fig 28a). Picture frame cracking similar to microindentation studies were also observed (Fig 28b). The load depth curves for two indentation depths (500 nm; 2000 nm) are shown in Fig 28c. The substrate effect was more prominent for larger indentation depths. The kink observed is due to pressure induced phase transformation of the crystalline Si substrate [23]. Crystalline Si substrates are studied for pressure-induced phase transformation under indentation at room temperature (RT) using a Berkovich tip nanoindentation. Raman spectroscopy, as a nondestructive tool, is used for the identification of the transformed phases [24]. A detailed explanation of the failure phenomenon occurring during nanoindentation can be found in [25, 26].

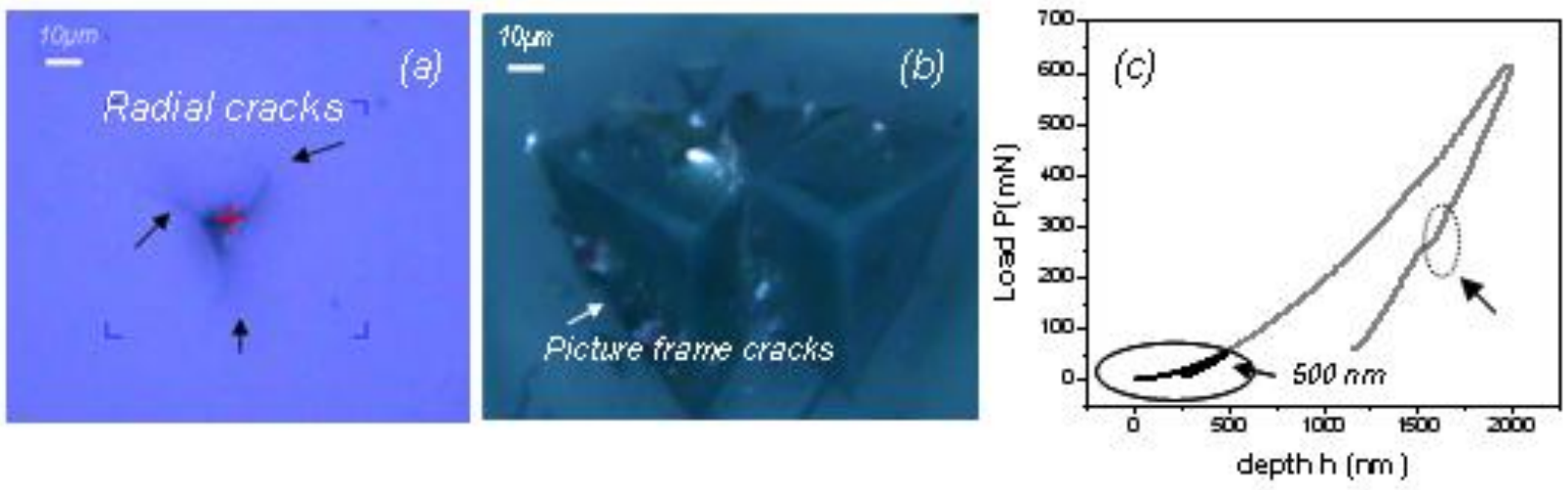

Fig 28: (a) Radial and (b) picture frame cracking occurring during nanoindentation (c) nanoindentation load depth curves of SiCN coatings deposited on Si substrates [17]

The load-depth curve obtained during nanoindentation gives the load which is required to penetrate a certain depth in the material and also provides the elastic and plastic work done during the indentation process. In our previous publication nanoindentation load depth curves were computationally generated and an attempt was made to throw light on some of the parameters during the loading and unloading process giving rise to different shapes of the curves [27]. In this article we provide a further detailed study of the indentation process keeping the indentation energy into consideration.

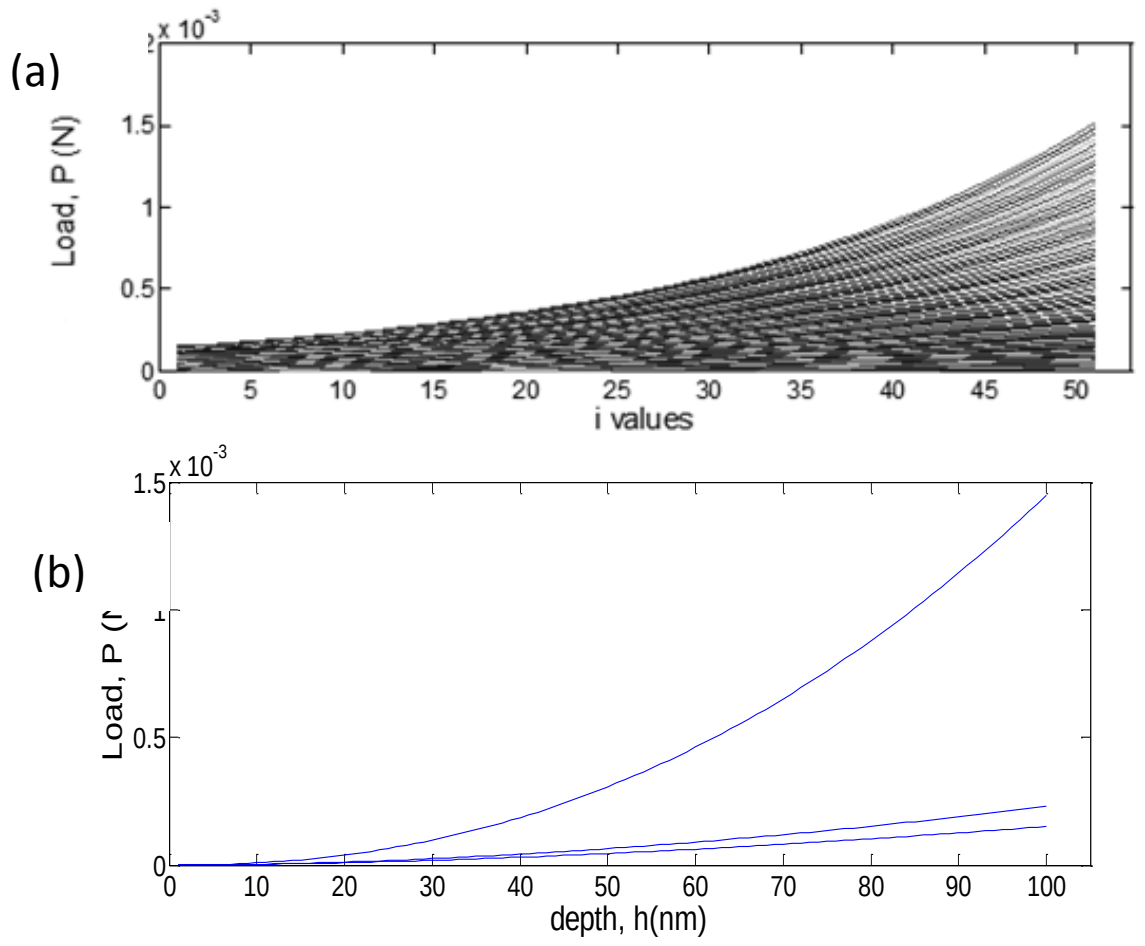

Fig 29 (a): Load vs. m (i) values and (b) Load vs. depth of Indentation

The relation P = k h $^2$ was used for the loading curve and P = k h $^2$ – k (h *max* - h) $^2$ for the unloading curve where P is the load and h is the depth of penetration. The value of the k was found to be 2.19e-4. The shape of the indenter is related to the parameters k, m, n and r which control the shape of the load depth curve; m , n, r being of powers of *k* ($k^n$), h during loading ($kh^m$) and (*h max – h*) during unloading ((h *max* – h) $^r$). For an indenter which has not gone any tip blunting, the parameters are taken as *m* = *2, n* = *1 and r* = *2*. However, the parameters vary with experimental conditions and we obtain ∫ [k $^n$ (h *max* – h) $^r$ ] d h as elastic energy, ∫ [k h $^m$ –k $^n$ (h *max* – h) $^r$ ] d h as plastic energy and ∫ k h $^m$ d h as total energy. In this computational study, the energy values were varied with the parameters. The parameter m is a reflection of changing contact geometry of the indenter as it penetrates [25]. The parameter n which also determines the curvature and slope of the curve is a reflection of the indenter tip sharpness. Lastly, the parameter r shows the elastic-plastic response of the material under test [27]. The parameter m has 51 values starting from 1.75 to 2.25 with an increment of 0.01.

The indentation load variation with depth for the different values is shown in fig 29(a) which is just an extension of the plot given in our previous publication[1]. The variation of load with m *(i=1, 10, 50)* values are shown in fig 29(b).The elastic energy was found to vary with *j i.e. n* value. However, it was constant for i or m value. The elastic energy distribution was same at different depth of penetration shown in fig 30(a). The plastic energy (EP) variation with *m (i)* is shown in fig 30(b). The total energy ET should theoretically be equal to the sum of elastic and plastic energy. However, it has been not been found to be following the same relation in every instance (Fig 30(c)). The difference in the value *dE = (EE+EP)-ET* indicates shear stress arising in the sample which is related to dislocation glide. The stress field around the dislocation encounters dislocation before the elastic to plastic transition [28]. A dislocation loop formed due to this is again of the same size as tip radius and hence dependent on m. Positive values of dE indicates compressive stress whereas negative values indicate tensile stress. So we can infer a tensile to compressive stress transition with increase in indentation.

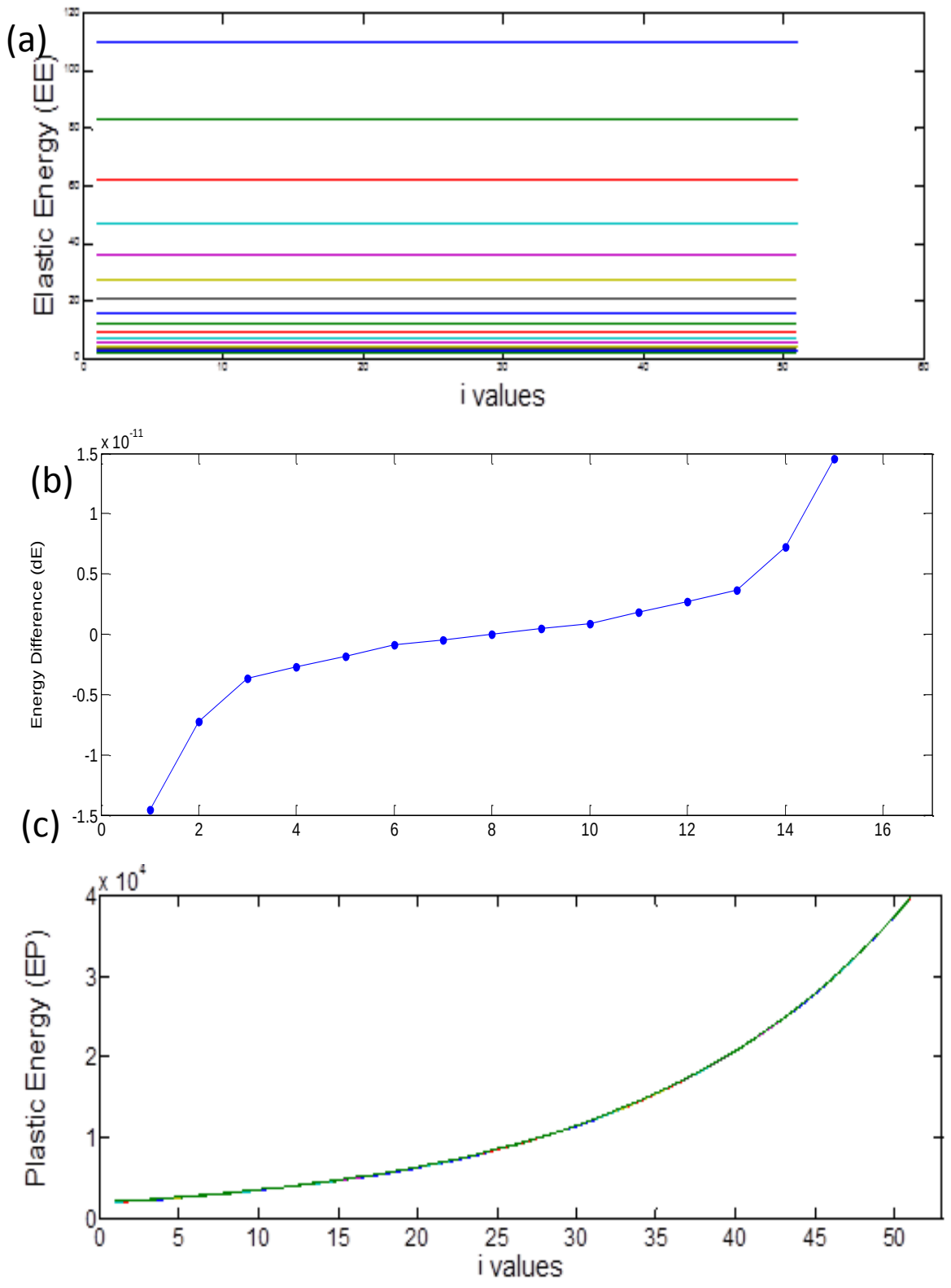

Fig 30: (a, b)Variation of elastic energy (EE), Plastic Energy (EP) with m(i) values (c) Energy difference of EP and EE with the total energy (ET)

## Acknowledgements

The author wishes to thank Dr. S.K. Mishra of CSIR-NML for experimental studies. A part of this article has been published in a conference proceeding [17] and the computational part on nanoindentation energy is also stored in archive of Researchgate.net [29]